\documentclass[acmsmall]{acmart}

\AtBeginDocument{%
  \providecommand\BibTeX{{%
    \normalfont B\kern-0.5em{\scshape i\kern-0.25em b}\kern-0.8em\TeX}}}

\usepackage{subcaption}

\usepackage{makecell}

\begin{document}

\title[Strategies and Designs to Facilitate Cross-Partisan Online Discussions]{`Walking Into a Fire Hoping You Don't Catch': Strategies and Designs to Facilitate Cross-Partisan Online Discussions}

\author{Ashwin Rajadesingan}
\affiliation{%
 \institution{University of Michigan, Ann Arbor}
}
\email{arajades@umich.edu}
\author{Carolyn Duran}
\affiliation{%
 \institution{University of Michigan, Ann Arbor}
}
\email{cduran@umich.edu}
\author{Paul Resnick}
\affiliation{%
 \institution{University of Michigan, Ann Arbor}
}
\email{presnick@umich.edu}
\author{Ceren Budak}
\affiliation{%
 \institution{University of Michigan, Ann Arbor}
}
\email{cbudak@umich.edu}


\setcopyright{acmlicensed}
\acmJournal{PACMHCI}
\acmYear{2021} \acmVolume{5} \acmNumber{CSCW2} \acmArticle{393} \acmMonth{10} \acmPrice{15.00}\acmDOI{10.1145/3479537}
\renewcommand{\shortauthors}{Rajadesingan et al.}
\begin{abstract}
While cross-partisan conversations are central to a vibrant democracy, these are hard conversations to have, especially in the United States amidst unprecedented levels of partisan animosity. Such interactions often devolve into name-calling and personal attacks. We report on a qualitative study of 17 US residents who have engaged with outpartisans on Reddit, to understand their expectations and the strategies they adopt in such interactions. We find that users have multiple, sometimes contradictory expectations of these conversations, ranging from deliberative discussions to entertainment and banter, which adds to the challenge of finding conversations they like. Through experience, users have refined multiple strategies to foster good cross-partisan engagement. Contrary to offline settings where knowing about the interlocutor can help manage disagreements, on Reddit, some users look to actively learn as little as possible about their outpartisan interlocutors for fear that such information may bias their interactions. Through design probes about hypothetical features intended to reduce partisan hostility, we find that users are actually open to knowing certain kinds of information about their interlocutors, such as non-political subreddits that they both participate in, and to having that information made visible to their interlocutors. However, making other information visible, such as the other subreddits that they participate in or previous comments they posted, though potentially humanizing, raises concerns around privacy and misuse of that information for personal attacks.
\end{abstract}



\maketitle

\section{Introduction}

Casual political conversations through which individuals build their identities, explore alternate perspectives and form considered opinions are key to a deliberative democracy \cite{Kim_Kim_2008}. Many of these political interactions take place in social media where users discuss politics among other topics with friends, acquaintances and often, strangers. Although online political interactions are associated with some positive outcomes such as increased civic participation \cite{shah2005information}, they are often unpleasant experiences; about 70\% of social media users report feeling stressed and frustrated when discussing politics with others on social media that they disagree with \cite{Anderson_2020}. Worryingly, the tone of political discussions online tends to be angrier, less civil and less respectful than offline conversations \cite{Duggan_Smith_2016}.  

A major factor contributing to hostility in both online and offline political discussions is the heightened levels of affective polarization that we observe today, a tendency of partisans to view opposing partisans negatively and copartisans positively \cite{iyengar2012affect}. Increasingly, rank-and-file Republicans and Democrats view each other as selfish, hypocritical and close-minded \cite{iyengar2019origins}. This increased outparty animosity is explained by Social Identity Theory which argues that by merely categorizing individuals into groups (here, Republicans and Democrats), group identities are activated, creating an ‘us’ versus ‘them’ group dynamic \cite{tajfel1979integrative}. Importantly, unlike protected attributes such as race where group-related behaviors are moderated by strong social norms and laws against discrimination, no such norms temper partisan hostility \cite{iyengar2019origins}. Thus, platform designers must account for and mitigate the deleterious effects of partisan identity when building systems that facilitate cross-partisan discourse.

Most prior research on improving cross-partisan discourse has predominantly aimed at addressing partisan bias in information consumption to burst filter bubbles \cite{pariser2011filter,munson2013encouraging,nelimarkka2019re}, with little emphasis on mitigating the role of partisan identity during interactions. In this work, we aim to reduce partisan prejudice by designing interfaces showcasing user information to promote cross-categorization and decategorization---two strategies adopted from social psychology research on inter-group conflict. Cross-categorization \textit{increases awareness of cross-cutting identities with members of the outgroup} ~\cite{brewer2000reducing}. Decategorization \textit{increases awareness of the distinctiveness of individual members of the ingroup and outgroup}~\cite{brewer1984beyond}. We conduct a qualitative study using semi-structured interviews (i) to first understand the expectations, concerns and strategies of users who engage in cross-partisan interactions and (ii) to seek feedback on designs and evaluate the types of information that can facilitate better cross-partisan discussions. We focus our analysis on Reddit, a popular social networking discussion site which hosts hundreds of political discussion communities (subreddits).

Our interviews reveal complex, and at times contradicting, motivations for participation in online cross-partisan talk, where participants look for serious deliberation but also entertainment and banter in political discussions. Participants also highlighted varied concerns with engaging in cross-partisan discourse. As one participant succinctly put it, cross-partisan talk can sometimes feel like ``walking into a fire hoping you don't catch'', requiring refined strategies to increase the odds of having compelling discussions. However, our designs to decategorize and cross-categorize users produced mixed effects. While participants expressed strong support for the cross-categorization inspired ``shared subreddit'' component, they---especially women and minorities---expressed that the extra user information provided by other components, while potentially humanizing, increased scrutiny on their profiles and would likely be used to attack them or derail discussions. We discuss the implications of these findings and detail the design challenges and opportunities to improve online cross-partisan discourse.

\section{Related Work}
\subsection{Partisan identity in online deliberation}

Normative theories of deliberation largely stem from J\"{u}rgen Habermas' conception of the public sphere, where citizens engage in rational-critical argumentation to form public opinion \cite{habermas1991structural}. The presupposed conditions central to such argumentation such as inclusion, discursive equality, ideal role-taking (impartiality and reciprocity) and absence of coercive power have been conceptualized as ideals of deliberation by deliberative theorists \cite{bachtiger2018deliberative}. While these ideals aim to ensure that individuals are swayed only by the best of arguments, in practice, empirical research reveals how partisan identities play a consequential role in how people engage with outpartisans and their arguments \cite{hendriks2007turning}. Motivated to maintain their party's positive distinctiveness and advance group status, partisans engage in inparty favoritism and outparty animosity \cite{huddy2017political}. This results in increased partisan hostility which we review below.

Partisan hostility typically manifests in the form of incivility and abuse targeted at outparty supporters. Partisans are more willing to denigrate outpartisans while judging incivility expressed by outpartisans more strongly than incivility by copartisans~\cite{rains2017incivility}. Exposure to copartisan attacks on outpartisans encourages copy-cat attacks while attacks by outpartisans result in stronger retaliation ~\cite{gervais2015incivility}. Further, this partisan hostility is often favored, even in highly moderated online discussion spaces; \textcolor{black}{studying the heavy moderated New York Times comments section, researchers observed that uncivil partisan comments received more ``recommendations'' (similar to upvotes) from users than comments that contained only uncivil language or only partisan language \cite{muddiman2017news}.}  Worryingly, exposure to partisan ad hominem criticism in news comments, which are exceedingly common online, result in more prejudiced attitudes towards outpartisans further exacerbating affective polarization \cite{suhay2018polarizing}.

These group-motivated behaviors may even be exacerbated in online spaces. \textcolor{black}{The Social Identity Theory of Deindividuation Effects (SIDE) posits that visual anonymity afforded by online platforms because of the lack of visible individuating information about group members increases the salience of group identities and adherence to group normative behavior \cite{Reicher_Spears_Postmes_1995}.} Following self-categorization theory, in the presence of an accessible group identity, individuals become depersonalized and view themselves and others less as individuals having distinct personalities but instead as interchangeable group members \cite{turner1987rediscovering}. Although commonly used to explain behavior in intragroup contexts, similar group dynamics have been observed in intergroup contexts as well. Through a series of experiments in intergroup online settings where participants were anonymous except for their group membership labels, Postmes et al.~\cite{postmes2002intergroup,postmes2005intergroup} showed that the depersonalization predicted by the SIDE model increased the relative salience of group boundaries and led to stereotyped perception of the outgroup. An important condition for observing depersonalization is that group identity must be accessible and salient during intergroup interactions. In the context of online political discussions, prior research overwhelmingly points to the salience of partisan identity in these interactions \cite{settle2018frenemies}. \textcolor{black}{Moreover, in many subreddits such as r/AskTrumpSupporters and r/AskALiberal, users, who already have very little individuating information about them (such as a profile picture, bio-sketch etc.) are also required to use a tag to identify themselves as a `Trump Supporter' or a `Progressive', setting up the ideal conditions for group-motivated behavior where a lack of individuating information is coupled with the presence of group cues.}

\subsection{Cross-categorization and decategorization to reduce partisan hostility}
Given that categorization into partisan groups forms the basis for partisan hostility, we review two social psychology approaches aimed at changing individuals' level of categorization: cross-categorization \cite{brewer2000reducing}  and decategorization \cite{brewer1984beyond}.
These strategies rely on the fact that individuals have multiple social identities apart from their partisan identities which may be activated to affect interaction dynamics \cite{hogg1987intergroup}.

\subsubsection{Cross-categorization} \label{Cross-categorization} Cross-categorization aims to make individuals of a group aware that they share membership in another dimension with individuals of the outgroup \cite{brewer2000reducing}. Revealing overlapping or shared group memberships makes social categorization more complex and reduces bias by increasing awareness of multiple subgroups within the outgroup \cite{crisp2006crossed}. Further, by making cross-cutting identities more salient, assimilation effects of the cross-cutting identity tend to offset the discriminatory nature of the partisan identity. Studying how other identities interact with partisan identity, Mason \cite{mason2016cross} observed ``a cross-cutting calm'', individuals with cross-cutting identities (for example, secular Republicans and evangelical Democrats) significantly reduced angry responses to party threats, exhibiting anger at even lower rates than weak partisans. Recently, testing the effects of shared non-political identities on partisan hostility, Levendusky experimentally found that individuals exhibited significantly higher warmth (by over 20\%) towards outpartisans when they were identified as supporting the same football team compared to no team identification (\cite{levendusky2020our}, Chapter 3). Based on these findings, we design an interface that surfaces ``shared subreddits'', users' shared membership in other nonpolitical communities, during their interaction to reduce hostility stemming from partisan identity. By explicitly highlighting shared group membership, we alert the user to the presence of ``calming'' cross-cutting identities.

\subsubsection{Decategorization} \label{Decategorization} 
Decategorization is aimed at increasing the salience of intragroup variability by highlighting the distinctiveness of individual members \cite{crisp2007multiple}. By exposing individuals to information about multiple other group memberships of outgroup members, individuals are nudged to differentiate outgroup members from the outgroup stereotype. Thus, by providing a more complex view of each outgroup member, individuals can evaluate them based on their personal merit rather than their stereotypical group memberships \cite{brewer1984beyond}. In politics, research suggests that people consistently stereotype outpartisans as being politically engaged extreme ideologues when no other information is provided about them which exacerbates outpartisan hostility \cite{druckman_misestimating}; when outpartisans were instead described as talking politics rarely and being ideologically moderate (who in reality is the modal outpartisan), outpartisans were evaluated more positively \cite{druckman_misestimating,klar2018affective}. Similarly, participants evaluated outpartisans who were less interested in politics more positively in a hypothetical roommate selection experiment \cite{shafranek2019political}. These findings suggest that providing information contextualizing the extent of users' political versus non-political attachments may help reduce partisan hostility. Thus, in addition to highlighting shared subreddits in our design, we also provide non-political individuating information about outpartisans in the form of ``active subreddits'', non-political subreddits that the interlocutor has recently participated in. By explicitly highlighting the other group memberships, we aim to decategorize the user as solely a member of their partisan group, instead we showcase the user as a distinctive individual with varied interests and identities, unrelated to their political leanings.

Another intervention closely related to this work is the intergroup contact hypothesis. The contact hypothesis suggests that interpersonal interactions between outgroup members under certain conditions: equal status, common goals, cooperative and institutional support will reduce intergroup prejudice \cite{pettigrew1998intergroup}. However, as Wojcieszak and Warner \cite{wojcieszak2020can} note, the intergroup contact hypothesis has not been extensively tested in the context of partisanship. While intergroup contact is central to this study, we aim to facilitate positive intergroup contact by reducing partisan bias, whereas studies testing the intergroup contact hypothesis examine the effects of intergroup contact on reducing partisan bias.

\subsection{Designing for online deliberation}

Researchers aiming to improve political deliberation have typically focused on two aspects: diversifying information consumption \cite{munson2013encouraging,munson2010presenting,Nelimarkka_Rancy_Grygiel_Semaan_2019} and facilitating deliberative interactions  \cite{Kriplean_Morgan_Freelon_Borning_Bennett_2012,Kriplean_Toomim_Morgan_Borning_Ko_2012}. As this work primarily concerns the latter, we review in detail the innovative interface designs that facilitate quality deliberation while reducing hostility in discussions.
Early work on online deliberation centered around highly structured interactions mapping information into facts, positions, arguments and relationships between them \cite{shum2008cohere}. In practice, these formal systems erected high barriers to usage as they required training to help users navigate complex predetermined interaction structures and argumentation schema \cite{shipman1999formality}. Over the past decade, researchers have aimed to facilitate high-quality deliberation while reducing such impediments, focusing on design considerations that center active contribution, navigability, usability, quality content and adoption \cite{towne2012design}. Kriplean et al. \cite{Kriplean_Morgan_Freelon_Borning_Bennett_2012} introduced ConsiderIt, a system that facilitates reflection of others’ perspectives by allowing users to form their own pro/con list on a particular topic by also including pro/con points contributed by others. Kriplean and colleagues \cite{Kriplean_Toomim_Morgan_Borning_Ko_2012} also built Reflect, a commenting system that makes active listening the normative behavior for users of the system by including a small listening text box along with the comment for users to succinctly summarize the original comment. Another system, OpinionSpace \cite{faridani2010opinion} maps users to points in a 2-D space based on their responses to five general value-based questions (answers to these questions map to either liberal or conservative leaning opinions), with the distance between the points representing the similarity between user answers to the question set. When a user clicks on a point, they can rate how much they agree and respect a comment posted by the user corresponding to the point. These systems all aim to make conversations more reflective. Alternately, finding that users often used multiple social media platforms,
Semaan et al. built Poli \cite{Semaan_Faucett_Robertson_Maruyama_Douglas_2015,semaan2014social}, an integrated political deliberation environment that aggregates multiple social media.

\subsubsection{Managing hostility in online deliberation}
Hostility stemming from interactions have been typically handled in two ways: (i) by structuring interactions to reduce direct contact (for example, ConsiderIt uses pro/con lists instead of facilitating back and forth interactions between users) and (ii) by removing or sanctioning problematic content, users or even entire communities \cite{jhaver2019did,seering2017shaping,chandrasekharan2017you}. More recently, researchers have aimed to design interfaces to proactively reduce hostility. Seering et al. \cite{Seering_Fang_Damasco_Chen_Sun_Kaufman_2019} designed psychologically embedded CAPTCHAs to prime users (just before replying) to trigger positive emotions that increased positivity, analytical complexity and interpersonal connectedness even in cross-partisan situations. Grevet et al. \cite{grevet2014managing} studying how weak ties manage political differences on Facebook, recommend another proactive approach; they suggest that ``making common  ground  visible (i.e., highlighting past interactions and shared  interests) during contentious  discussions could alleviate in-the-moment tension.'' This lends further support to our design choice to highlight shared subreddits during interactions. Although Reddit users are unlikely to know each other unlike Facebook, we expect that showing shared non-political group memberships will likely still have an effect of alleviating tension. \textcolor{black}{Somewhat paradoxically, many of Reddit's design choices (e.g. up/down voting mechanisms) and participation cultures (e.g. circlejerking\footnote{Circlejerking defined by Allison et al. \cite{allison2020communal} as "a slang term referring to the mutual appeal to and gratification of shared interests and tastes within a community"}) which contribute to the insularity of the subreddits may actually strengthen the effects of shared memberships in these communities by increasing users' bonds with other community members \cite{allison2020communal}.}

\subsubsection{Managing partisan identities in online deliberation}
Despite the prominence of partisanship in political interactions, most systems or designs (barring a few notable exceptions such as ConsiderIt and OpinionSpace) do not specifically address the prevailing group dynamics in these interactions. ConsiderIt \cite{Kriplean_Morgan_Freelon_Borning_Bennett_2012} takes the deliberate strategy of providing no information about users beyond their names to ``not provide group cues to activate political identity''. OpinionSpace \cite{faridani2010opinion} takes the opposite approach of displaying users according to their answers to the values question set as described earlier. It takes advantage of the fact that liberal and conservative users often have similar answers to the values question set, resulting in closely spaced points in the 2-D space, contrary to expectations of seeing them on opposite ends of the space. This disrupts users' binary mental models and ``conveys that the range of opinions do not fall along a single axis  and that they are far more diverse.'' With both shared and active subreddits, we build on OpinionSpace's underlying principle that revealing information about users would show that users are not as divided as they are projected to be. By showcasing non-political group memberships, users are presented with a more complicated picture about their interlocutors which we expect will disrupt the `us' vs `them' partisan group dynamics.

\subsubsection{Exposing user information in online deliberation}
A significant concern with online deliberative systems is that interactions are often between users who know nothing about each other, leading to concerns about trust and credibility of information exchanged \cite{wang2020influence}. For example, Kriplean and colleagues on evaluating ConsiderIt noted that ``almost immediately after raising the issue of trust, user study participants would comment that they wanted to know more about the point author.'' However, as discussed above, they do not include user details to prevent priming partisan identity. In contrast, our design choice to show non-political group activity details to reduce partisan identity salience may also help to increase trust by providing individuating information. For example, Tanis et al. \cite{tanis2005social} found that, as predicted by the SIDE model, revealing individuating information about an anonymous outgroup member online increased interpersonal trustworthiness as the member is seen less as an outgroup member and more as an individual. However, revealing information about group memberships comes with multiple concerns. Firstly, it raises concerns about inadvertently revealing sensitive private attributes \cite{zheleva2009join}. Secondly, revealing this information may result in an asymmetrical disclosure, where one party knows information about the other but not vice-versa. Studies, albeit on dating practices, show that even when this information is obtained from public Facebook profiles, it is typically considered deceptive and norm violating to use it \cite{hancock2008know}. Finally, this information initiates a form of `context collapse' \cite{marwick2011tweet}. On Reddit, usually user activity in one subreddit is not directly visible in another subreddit allowing users to relatively freely participate in subreddits related to unpopular or stigmatized topics without it affecting their other activities (although throwaway accounts are still common) \cite{de2014mental}. Thus, disclosing this participation information can cause real harm and harassment, especially given Reddit's known toxic participatory cultures \cite{massanari2017gamergate}. Therefore, we carefully evaluate if and when users consent to share their activity details with others.

\section{Research Methods}
\subsection{Research Context}
We conducted this study on Reddit users in the lead-up to the 2020 U.S. Presidential elections. Reddit is a popular social networking platform comprising of hundreds of thousands of subcommunities called subreddits. Each subreddit is centered around a topic and independently run by volunteer moderators. Although there are some commonalities, the norms and rules enforced in these subreddits may also vary significantly \cite{chandrasekharan2018internet,fiesler2018reddit}. For example, r/NeutralPolitics and r/moderatepolitics both host cross-partisan discussions but vary in how the discussions are conducted. While the former does not allow ``bare expressions of opinion'' and requires claims to be backed by sources, the latter has no such restrictions. Users interact with each other in these subreddits through a threaded comment system that allows users to directly reply to each other. This allows for prolonged interactions between pairs of users. Comments accumulate points (called karma) through up/down votes by other users which affect their visibility. Users accumulate karma points as well, which is the sum of their comments' karma points. A similar mechanism applies to the top-level posts in the subreddits called ``submissions''. Many cross-partisan interactions take place usually in relatively non-partisan subreddits such as r/PoliticalDiscussion, question-answer subreddits such as r/AskTrumpSupporters, ideological subreddits such as r/neoliberal and occasionally in partisan subreddits such as r/politics. As an indicator of the levels of partisan animosity prevalent on Reddit, many large political subreddits such as r/The\_Donald and r/ChapoTrapHouse were banned for inciting hate just a few days before our first interview. It is in this context that we studied the strategies that users engage in cross-partisan discussions and the potential effectiveness of our designs in facilitating quality discourse.
\begin{table}
\small
\begin{tabular}{ |c|c|c|l|l|l|c| }
 \hline
Participant & Recruitment & Age & Gender & Ethnicity & \makecell{Political \\Orientation} & \makecell{Years on\\ Reddit}\\
 \hline
P01 & PM & 37 & Male & White & Left & 10\\
 \hline
P02 & PM &19 & Male & East Asian & Left & 1\\
 \hline
P03 & PM &23 & Male & White & Right & 3\\
 \hline
P04 & PM &35 & Male & White & Left & 7\\
 \hline
P05 & PM &36 & Male & Caucasian & Ind./Right-leaning & 10\\
 \hline
P06 & Univ. &20 & Male & Caucasian & Right & 5\\
 \hline
P07 & PM &21 & Male & White & Right & 3\\
 \hline
P08 & Univ. & 25 & Female &  Chinese-American  & Left & 6\\
 \hline
P09 & Univ. & 24 & Male &  \makecell[l]{Hispanic / Latino \\and White} & Left & 7\\
 \hline
P10 & Univ. & 28 & Male & \makecell[l]{Middle Eastern / \\Southwest Asian} & Ind./Right-leaning & 15\\
 \hline
P11 & Post &- & Female & Black & Left & 2\\
 \hline
P12 & PM & 37 & Male &  White & Right & 5.5\\
 \hline
P13 & PM & 48 & Female & Jewish & Left & 7\\
 \hline
P14 & Univ. & 23 & \makecell[l]{Nonbinary / \\ Genderqueer} & (Southeastern) Asian & Left & 8\\
 \hline
P16 & PM & 62 & Female & \makecell[l]{Caucasian and \\Native American} &
\makecell[l]{Right / \\NeverTrump} & 1.5\\
 \hline
P17 & Post & 33 & Male & Black & Left & 1.5\\
 \hline
P18 & Post & 22 & Female & Caucasian & Left & 5 \\
 \hline
    
\end{tabular}
\caption{Demographic details of participants. We report participant responses to a short open-ended demographic survey as submitted by them. P11 did not provide age details. Recruitment channels are PM (Reddit private message), Univ. (university mailing lists) and Post (post on subreddits).}
\medskip
\small
\label{table:demographics}
\end{table}
\subsection{Participants and Recruitment}
The participants of this study are United States residents who actively use Reddit to have cross-partisan political discussions. Participants were recruited through Reddit private messages, recruitment posts on subreddits such as r/PaidStudies and multiple university mailing lists. First, we tried recruiting by sending private messages on Reddit to users inviting them to participate in the study from a Reddit account created for this purpose; we did not get any responses. Speculating that the lack of response was due to the account being new and not trusted, we sent recruiting messages through the first author's personal account which was much older, had more karma points and  detailed history. This approach was more successful, 9 out of 83 ($>10\%$) users to whom we reached out agreed to participate in the study. We sent recruitment messages to users who actively engaged with opposing partisans in political subreddits such as r/politics, r/AskTrumpSupporters and r/moderatepolitics. However, this approach appeared to predominantly recruit White males, likely due to privacy and safety concerns. Therefore, we turned to two other channels: university mailing lists and subreddits such as r/PaidStudies. These are both popular recruiting avenues for academic research where we could more easily identify ourselves as university researchers to establish trust. We were able to recruit a more diverse set of participants using these approaches. The interviews were conducted from July to September of 2020. In total, we conducted interviews with 18 participants. For this paper, we exclude P15 who in her interview explained that she only lurked on political subreddits and did not participate in them. Participants were required to be (i) US residents, (ii) 18 years or older and (iii) must have participated in cross-partisan discussions to be eligible for the study. Each participant was paid with a \$20 Amazon gift card as compensation for their participation in the study.

Table \ref{table:demographics} lists the demographic details of the participants.\footnote{We report participant responses as is from a short open-ended demographic survey.} \textcolor{black}{11 of the participants identified as male, 5 as female and 1 as nonbinary/genderqueer.} Participants ranged from 19 to 62 years of age, with most participants in their early twenties. They skewed mostly young, white, and male, paralleling the general demographics of Reddit users \footnote{www.journalism.org/2016/02/25/reddit-news-users-more-likely-to-be-male-young-and-digital-in-their-news-preferences/}. \textcolor{black}{10 of the participants were left-leaning, 5 were right-leaning and 2 were right-leaning independents.} We interviewed participants in different occupations such as software programmers, university administration staff, high school teachers, census workers, undergraduate and graduate students. P12 is also a moderator of a political subreddit. Participants' experience on Reddit range from 1-15 years with a median of 6 years of involvement, and many spent months lurking before creating their account.

\subsection{Data Collection}
The interviews were conducted by the first and second authors. Almost all participants were interviewed using video conferencing software (except P16 with whom we conducted a telephonic interview and narrated the designs instead). The audio was recorded after obtaining informed consent and later transcribed. The median duration of the interviews was 55 minutes. Each interview consisted of two parts: a semi-structured interview (around 40 minutes) and a design probe interview (around 15 minutes). From the semi-structured interviews, we obtained rich and detailed information on their motivations, positive and negative discussion experiences, and strategies they use to participate in these discussions. In the design probe part of the interview, we shared 2-3 designs based on decategorization and cross-categorization strategies on screen and after a brief explanation of the probe, we asked for their feedback and reactions to the probe. We also specifically probed for concerns they may have had about using the interface and about others using this interface when interacting with them.

\subsection{Data Analysis}
Each interview was transcribed using otter.ai before manual revisions and corrections by the first and second authors. The interviews were coded using a grounded theory approach \cite{charmaz2006constructing} consisting of both open and axial coding using NVivo software. The first and second authors independently coded the interviews (12 and 5 interviews respectively) using open coding. These codes were then combined into higher level categories using an axial coding process. The two authors met multiple times to discuss and combine these categories, and identified emerging themes around (i) motivations for participating in cross-partisan discussions, (ii) qualities of political discussions, (iii) proactive and reactive strategies adopted by participants to have good discussions, (iv) folk theories of why cross-partisan discussions are difficult to sustain, and (v) humanizing effects of the design probes and concerns around misuse. Through the course of interviews, we held weekly meetings with the research team to discuss the feedback from interviews about the designs, allowing us to incorporate minor modifications to the design probes detailed in Section \ref{designprobes}.

\subsection{Design probes} \label{designprobes}

\begin{figure}
  \begin{subfigure}{\linewidth}
   \centering
\includegraphics[width=0.80\textwidth]{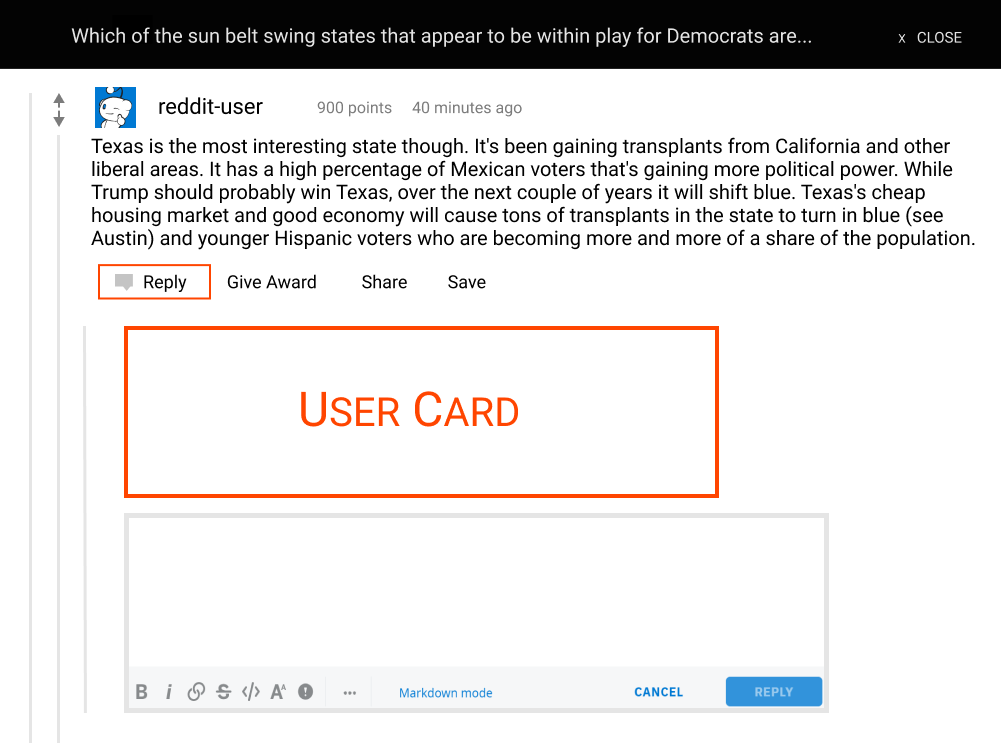}
    \caption{Example of a comment on the Reddit interface with our user card. The user card would appear when users click the reply button to type a reply.}
    \label{interface}
  \end{subfigure}\par\medskip
  \begin{subfigure}{\linewidth}
      \centering
    \includegraphics[width=0.80\textwidth]{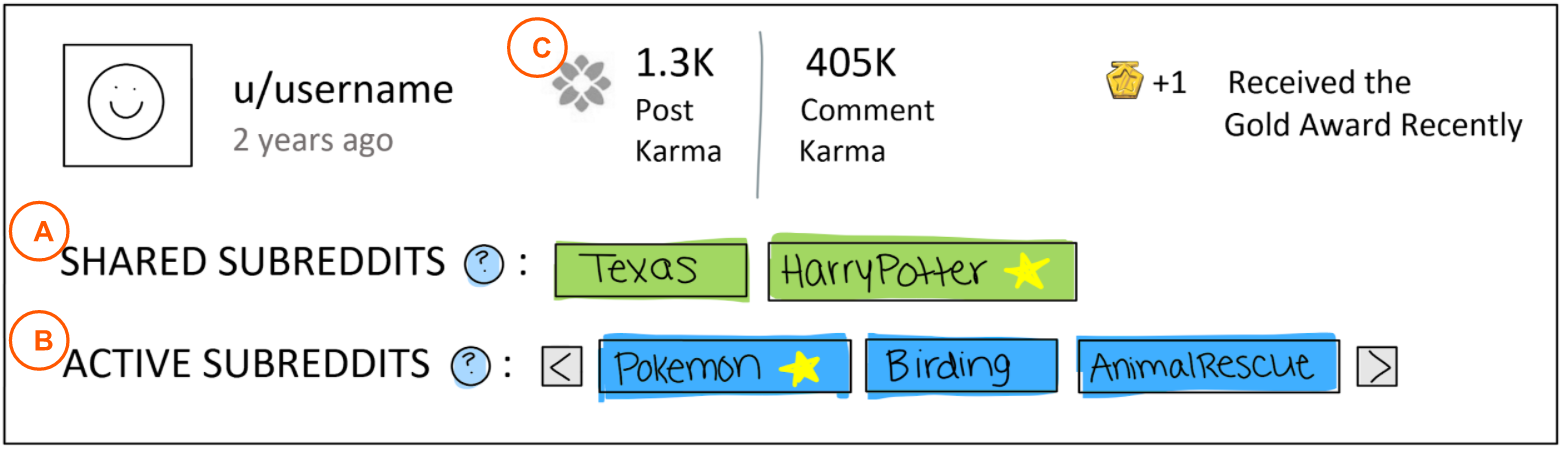}
    \caption{Design A: The user card shows active and shared subreddits as well as karma points and awards.}
    \label{usercard1}
  \end{subfigure}\par\medskip
  \begin{subfigure}{\linewidth}
   \centering
      \includegraphics[width=0.80\textwidth]{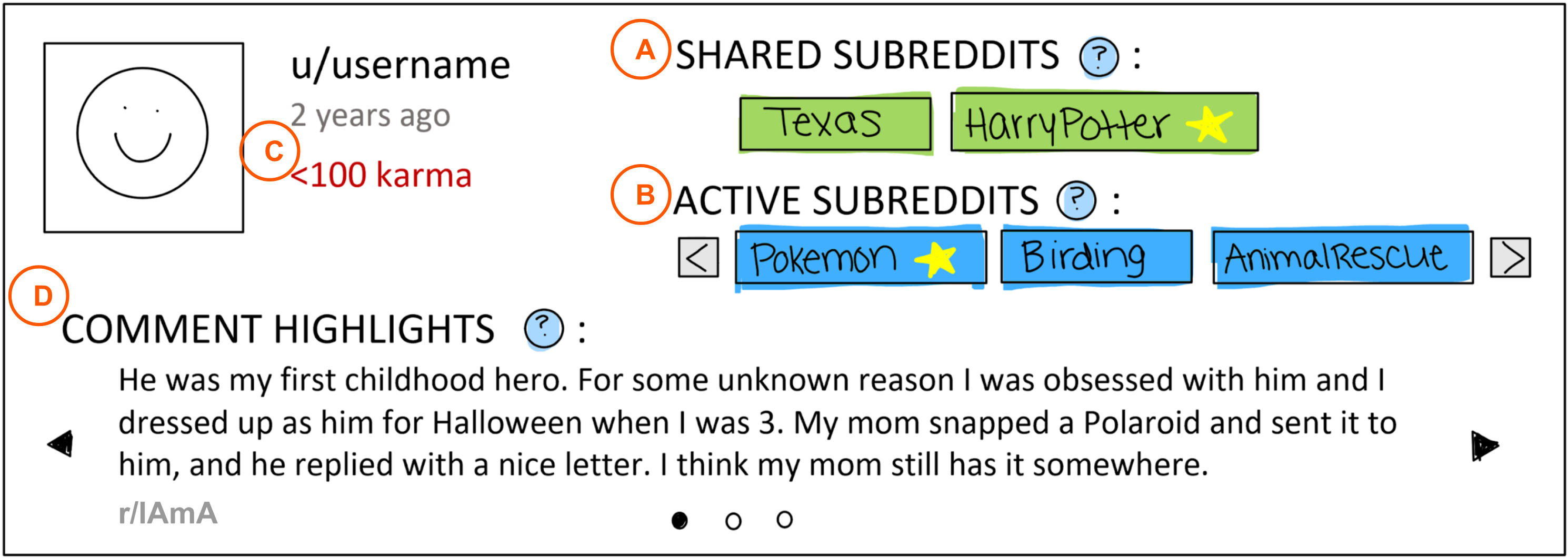}
    \caption{Design B: In addition to components in design A, the user card shows comment highlights.}
    \label{usercard2}
  \end{subfigure}
  \caption{User card designs}
\end{figure}

Currently, the Reddit interface, as shown in Figure \ref{interface} (the interface excluding the user card), does not directly provide any information about the interlocutor. Users need to hover over the username to obtain basic profile attributes such as time since joining Reddit and total karma points. To view past comments or other subreddits their interlocutor has participated in, the user has to go to the interlocutor's profile page by clicking on the profile icon. Through our design probes \cite{gaver1999design}, we explore alternate versions of the Reddit interface where the user has access to additional information which is expected to decategorize or cross-categorize their interlocutor. By visually showing designs containing this extra information, as opposed to asking participants to imagine such a possibility, we provide a realistic representation of this information on which to base their opinions. The aim of the designs are two-fold: (i) To understand how participants perceived the impact of the extra information on their conversations and (ii) To explore different designs based on participant feedback to build a functional browser extension. Below, we detail each component of the user card which is intended to show up when users click the `reply' button to reply to another user's comment (as shown in Figure \ref{interface}).

\subsubsection{(A) Shared subreddits} \label{shared_sub}
This component shows the list of non-political subreddits that both the participant and their interlocutor have recently participated in. By explicitly highlighting shared group memberships, we alert the user to the presence of cross-cutting identities which is found to have a calming effect on partisan hostility as described in Section \ref{Cross-categorization}. The subreddits were be ordered such that smaller subreddits were shown first since group size is negatively associated with affinity towards the group in online communities \cite{kraut2020makes}.

\paragraph{Feasibility Analysis} \label{feas}
Displaying shared subreddit memberships to interlocutors will be beneficial only if they actually share subreddit memberships. This concern is especially significant now as political science research suggests conservatives and liberals on average make different choices on even non-political decisions such as coffee choice and fast food consumption \cite{dellaposta2015liberals}. This may result in few common subreddit memberships between outpartisans. Therefore, to evaluate the reach and thereby effectiveness, we estimate the prevalence of shared non-political group memberships among users who engage in cross-partisan discussions on Reddit using publicly available data \cite{baumgartner2020pushshift}.

First, using a simple heuristic from prior work on Reddit \cite{rajadesingan2021political}, we identify users who are left or right-leaning based on their activity in left and right-leaning subreddits. First, we identify  r/politics, r/Liberal, r/progressive as left-leaning and  r/TheDonald,  r/Conservative,  r/Republican as right-leaning subreddits. Then, we classify users as left-leaning if (i) they comment in more left-leaning than right-leaning subreddits (ii) the mean karma points of their comments in left-leaning subreddits is higher than their score in right-leaning ones and (iii) their mean karma score in left leaning subreddits is greater than 1. Likewise, we identify right-leaning users. \footnote{We classify 1,223,229 users as left leaning and 367,363 users as right leaning. We cannot identify the political leanings of other users using this approach.} Then, using these user classifications, we identify all distinct copartisan and cross-partisan interlocutor pairs in 277 political subreddits (previously identified by \cite{rajadesingan2020quick}). For each pair, we identify if they both participated in a common subreddit within the last 3 months, while excluding the 277 political subreddits and the default subreddits\footnote{Until June 2017, Reddit users were automatically subscribed to these subreddit which are amongst the largest on the site.} from consideration. We find that, in an average subreddit, 44.26\% and 51.94\% of all cross-partisan and copartisan discussion pairs share at least one common non-political subreddit. These percentages are encouraging because (i) in an political average subreddit, about half of all discussion pairs share a non-political subreddit indicating that showing shared subreddits is a viable option for a sizable population of interactions, (ii) \textcolor{black}{the difference between copartisan and cross-partisan percentages, although statistically significant, is small enough to suggest this difference may not significantly exacerbate outparty differences, although experimental work is required to estimate the effect of \textit{not} observing shared subreddits. However, these results also highlight an important limitation of this design: less than half of all cross-partisan interaction pairs can be shown the shared subreddits component.}

\subsubsection{(B) Active subreddits}
This component shows the list of non-political subreddits that the interlocutor has recently participated in, excluding the ``shared subreddits''. By explicitly highlighting the interlocutor's varied interests and identities based on their activity, we aim to reduce hostility through decategorization as described in Section  \ref{Decategorization}. Again, these subreddits were be ordered such that smaller subreddits were shown first.

\paragraph{Feasibility Analysis}
Displaying active (and not shared) subreddit memberships will be beneficial only if they actually participate in other subreddits. Redditors are known to create multiple throwaway accounts to provide added anonymity, especially when discussing contentious issues \cite{leavitt2015throwaway}. If users predominantly use throwaways when talking about politics, then there may be few other subreddits to display to interlocutors. Therefore, using a similar approach as earlier, we calculate among users who participate in political subreddits in a given month, the average number of nonpolitical subreddits they participated in the prior three months. We find that the left and right leaning users active in political subreddits in 2019, on average,  engage in about 23 and 20 subreddits respectively in the prior three months, providing evidence that this design is indeed feasible given current user behavior data.

\subsubsection{(C) Karma points and awards}
This component shows the karma points and awards earned by the interlocutor. Though unrelated to decatagorization or cross-categorization, karma points may have potential to improve conversations by providing an indicator of trust or reputation bestowed on the user by the Reddit community. This feature was designed based on feedback from ConsiderIt where their study participants expressed difficulty in evaluating the trustworthiness of claims put forth by other users about whom they knew nothing \cite{Kriplean_Morgan_Freelon_Borning_Bennett_2012}. Highlighting awards and karma points could present one way to highlight trust without giving away partisan cues about the user.

\subsubsection{(D) Comment highlights}
This component highlights top comments posted by users in non-political subreddits based on karma points. By providing examples of top non-political comments by the interlocutor, we aim to showcase their positive behavior in other subreddits indicating that they have multiple interests apart from their politics. Along with the active and shared subreddits, comment highlights provide deeper insights into not only where they participate but also how they do so in the other subreddits. A more discrete version of comment highlights is the star-shaped link in the active and shared subreddit boxes which links to a top comment (above 50 karma) posted by the user in that subreddit.

\paragraph{Design evolution}

First, we conducted interviews using one design (Design A, Figure \ref{usercard1}). During the first few interviews (P1-P3), participants suggested providing cues about not just \textit{where} users were active but also \textit{how} they behaved in those places. This feedback resulted in us evaluating additional designs that made visible more details such as ``comment highlights'' in Design B (Figure \ref{usercard2}). We showed Design A to all participants and Design B to P4-P18. We also developed other largely similar versions of these designs aimed at reducing the size of the user card by moving the placement of karma points, showing shared and active subreddits on the same row and linking to a highlighted comment rather than displaying full text. All designs featured shared and active subreddits, the primary focus of our research.

As a cautionary tale, we note that in our case, the initial mode of recruitment (private messages) resulted in mostly White/Caucasian male participants (P1-P7) who did not have major concerns about their information being made more visible in the user cards. However, as we interviewed a more diverse set of participants recruited through other channels later in the study, concerns about revealing information became clearer. We caution that designs such as ours that highlight user information needs to be carefully evaluated for their effects, especially on members of disadvantaged groups early in the design process.

\paragraph{Target demographics}

A significant advantage of these designs is that they are not explicitly political; users would simply see the non-political activity of other users. Therefore, an extension built using these designs can be marketed as a fun tool that helps users learn more about others, which we expect will help diversify the kinds of users who install the extension. By positioning it as a general-purpose fun tool, we anticipate that all users, not just the ones most motivated to improve their discussions, will use the extension. However, because of the nature of these designs, we expect it to be less effective on extreme partisans whose non-political subreddit membership is likely stereotypical. Further, the extension is not expected to reduce hostility expressed by individuals who are determined to be hostile, rather it is a subtle intervention aimed at users who engage in cross-partisan interactions in earnest.

\section{Findings}

\begin{figure}[h]
    \centering
    \includegraphics[width=1\textwidth]{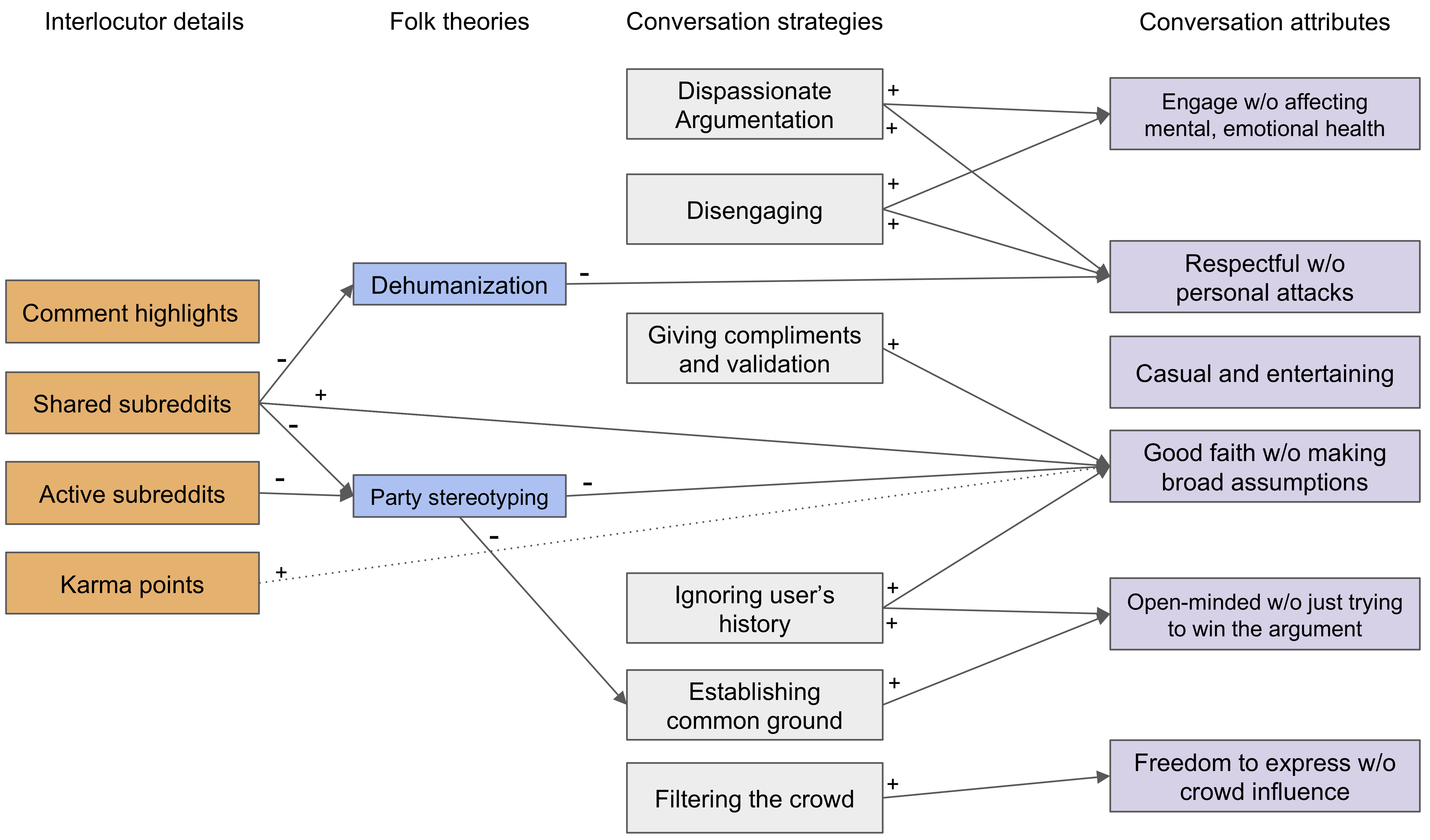}
    \caption{Summary of findings showing the relationships between good conversational attributes and user strategies employed during the conversation, folk theories and interlocutor details made available through design. The arrows indicate the directionality of the relationship between the entities, and the signs (+/-) indicate whether the relationships were positive or negative. For example, establishing common ground increases the odds of having open-minded discussions while an increase in dehumanization decreases the odds of having respectful interactions. The line from karma points to good faith is a dotted line since karma is a weak/basic indicator of good faith.}
    \label{fig:mediation_model}
\end{figure}

We organize our findings as follows: First, we detail the qualities that participants seek in a good cross-partisan political discussion (rightmost in Figure \ref{fig:mediation_model}). Next, we highlight strategies that participants adopt to improve the chances of experiencing these good qualities in their discussions (center-right, in grey). Then, we detail two folk theories--dehumanization and stereotyping--that participants attribute to the many bad conversations they have despite following these strategies (center-left, in blue). Finally, we explore how the user information embedded in our designs may help overcome dehumanization and stereotyping but may also lead to other concerns (leftmost). The (+) and (-) signs in Figure \ref{fig:mediation_model} indicate positive and negative relationships between the entities. For example, establishing common ground increases the odds of having open-minded discussions while an increase in dehumanization decreases the odds of having respectful interactions.

\subsection{What is a `good' cross-partisan political discussion?} \label{concerns}
When asked what they considered to be a good cross-partisan interaction, participants described two kinds of interactions: (i) serious deliberative discussions on political or policy issues and (ii) casual conversations for entertainment and banter. Interestingly, many participants reported engaging in both kinds of conversations depending on their mood or time constraints. We describe these conversations in detail below.

\subsubsection{Serious deliberative discussions}
Most participants expressed that they were looking for some form of serious deliberative discussion. Many of the specific attributes they looked for in such conversations directly mapped to the deliberative ideals of mutual respect, reasoned arguments and the freedom to express without coercion (such as crowd influence).

\paragraph{\underline{Respectful without name-calling and personal attacks}}
Most participants expressed that they aim for conversations to be polite and respectful without devolving into personal attacks.

\textit{The bad conversations start off right away antagonistically, you'll have like a Trump supporter or a liberal supporter like basically just start off by saying nasty ad hominem attacks about the other side, these are already like non-starters like you're not going to get anywhere.} - P01

\paragraph{\underline{Listening with an open mind without simply trying to win the argument}}
Most participants enter ed cross-partisan discussions without expectations of changing others' views. Instead, they looked for conversations where their interlocutors were simply open to acknowledging some of the issues that they had raised.

\textit{For instance, [pretend] you're pro-Trump. But I made a point that you can't find anything to disagree with about. Can you actually say that? You know, while I support Trump, you actually have a valid point on this particular issue. So, being willing to listen is to at least consider what the other person is saying, which does require listening, is huge.} - P16

However, participants indicated that many conversations are not open-minded exchanges of ideas but rather interlocutors simply trying to one-up each other. Thus, some participants do not even look for open-minded users, instead, they use the conversation to explore an issue. For example, P12 recounted instances where he argued with others ``for the sake of just understanding that idea''. Few of our participants actively looked to change others' views. Still, they reported that instances where they changed others' viewpoints were uncommon.
\paragraph{\underline{Good faith without making assumptions}}
Participants look forward to having good-faith conversations with others--conversations where everyone has good intentions, engages in earnest, and refrains from making assumptions of others.

\textit{[A good conversation needs] understanding that each person participating has experiences that you might not be able to relate to or like, language is really imperfect... understanding that like everyone is like trying to do right by their communities and families. So even if like we can't understand what those obligations look like, they are "good people"...} - P14

However, participants noted that in many of these conversations, people quickly make assumptions and judge others without giving them a chance to explain their beliefs.

\paragraph{\underline{Informative without unverifiable claims and party talking points}}
A prime motivation for almost all participants to engage in cross-partisan political discussions is to learn about opposing viewpoints and contribute alternate perspectives. Similar to Semaan et al.'s \cite{Semaan_Robertson_Douglas_Maruyama_2014} findings, most participants in our study explicitly acknowledged the effect of filter bubbles or echo chambers on their own beliefs. They stated that they actively try to engage in cross-partisan political discussions to gain alternate perspectives.

\textit{I enjoy talking with [conservatives] because they'll see an article and see it completely different than the way I see it. It's a curiosity for me... And it's good for me to know that they exist, and not just this little bubble that I'm in.} - P04

However, in many conversations, participants noticed that the interlocutors simply regurgitated party lines or spread debunked misinformation without researching on their own to understand the issue.

\textit{When the person is not willing to debate facts, when they start spewing basically talking points, talking points that are disputed, talking points that aren't related, talking points that don't make sense... then that's a pretty good indication it's not going anywhere.} - P17

\paragraph{\underline{Freedom to express without crowd influence}}

Almost all of our right-leaning participants and many left-leaning participants described how their comments are often heavily downvoted or heavily replied to by many users (dogpiled), which overwhelms them.

\textit{Reddit, a lot of it is primarily liberal. So it's like, if you want to come with any conservative opinion whatsoever, you're probably just going to get mobbed on and, you know, for every 100 people that mob on you, there might be five actual discussion points in there.} - P05

However, participants sometimes do take into account the feedback they receive from others, especially copartisan feedback. As P13 described, receiving downvotes or multiple replies does prompt her to reflect and question her positions on the issue.

\textit{People will reply vehemently... I'm really surprised when it's more than two people... It makes me wonder whether or not my position on that topic should be that position. Do I change my mind? No, not necessarily, but the thought is there and that's important too, because you have to constantly question your own thoughts.} - P13

\paragraph{\underline{Engaging without affecting mental and emotional health}}

Many specifically described the toll that some conversations had on mental health. Participants report that cross-partisan conversations often involve a lot of work and mental effort, the after-effects of which may continue to linger through the day. Even conversations that do not veer into name-calling or character attacks sometimes leave participants frustrated and exhausted.

\textit{The fear is that like, it’ll consume the whole day. I’ll be thinking about something politically... and I'll just keep talking about, thinking about like, something political and get worked up about it.} - P07

However, some participants are able to ignore views that they dislike, or quickly move on to lighter content to decompress, as P11 put it, ``continue to scroll to find a cute puppy''. Others, like P2, are resigned to the fact that being exposed to objectionable views is the cost of having a cross-partisan conversation.

\textit{But it's those little prejudices that people have that bother me, but while I do feel bad reading them, I don't think I am necessarily upset because of it. Because the main reason why I'm on there is for a political discussion.} - P02

\subsubsection{Casual and entertaining conversations}

Participants explained that they were on Reddit primarily to have fun and entertain themselves. They did not use Reddit for political discussion alone and most actively engaged in other relatively non-political communities such as r/DIY and r/Makeup. They saw their participation in political conversations as one among many other leisure activities they engaged in on Reddit. In fact, some participants explained that many of their political conversations were incidental and stemmed from casually browsing through their home feed. They were not actually trying to engage in the conversation deeply and would typically quickly comment and leave.

\textit{I kind of will just comment whatever, not really trying to seek out [conversation] because also, once you do get a productive conversation going, it takes a lot of energy, it can take a lot of time... I don't know if I have the stamina for it all the time.} - P17

Sometimes, participants engage in more casual political subreddits such as r/PoliticalHumor and r/PoliticalCompassMemes. However, many find discussions in mainstream discussion subreddits entertaining as well. P09 described how he uses Reddit most heavily when he is bored at work, and primarily looks for entertainment when participating in political discussions. Below, he described one such discussion where a conversation in r/politics devolved into a conspiracy theory.

\textit{One of the funniest ones I ever remember reading was a pizzagate-like thread of comments. It was just hilarious. Because that was a conspiracy theory and then lots of people branch off and like, I don't know, it was so entertaining... Like watching some people just put two and two together on things I could have, like, never thought twice. That's some high entertainment value.} - P09

Participants also mentioned that even within more serious discussions, someone may post a witty rejoinder or a funny meme which makes the tone of the conversation fun and casual. They also pointed to how some political subreddits have dedicated discord chat servers and occasional free talk threads, allowing users to have casual discussions unrelated to politics. In certain instances, participants indicated that normatively anti-social behavior such as trolling and making others angry were also fun activities to take part in on political subreddits.

\textit{If I am in a really bad mood, and I'm just out to, you know, troll up a storm, then what I think of as good [conversation] is when I get someone's goat, when I make them very viscerally angry, and they keep responding, and I can tell they don't want to respond, but they have to respond and that's when I've got them!} - P10

This poses a direct contradiction with one of the most commonly cited motivations---to have respectful deliberation. It is important to note that the same participants who seek entertainment in political discussions also participate in more deliberative political discussions. The kind of conversations in which participants choose to take part depends on a wide range of factors such as mood, time constraints and current events.

\paragraph{Frequency of ``good'' discussions}
Most participants reported that they participated in at least a few political discussions that they felt were good and satisfying. However, these occurrences were rare. Many long-time participants reflected that their conversations have turned angrier and devolved more into name-calling over the past two years. However, many participants characterize engaging with the other side as a form of civic duty, something that is difficult but needs to be done to deeply understand issues affecting the country. Therefore, to navigate these conversations and increase the odds of having a good interaction, participants have developed multiple strategies to select where, who and how to talk to cross-partisans which we detail in the following sections.

\subsection{Strategies adopted prior to engaging in cross-partisan discussions} \label{proactive}
\subsubsection{Choosing where to have the conversation}
\label{customizing}
Many participants reported taking part in multiple political subreddits and carefully curating the subreddits that they subscribed to, considering the quality of discussion, member composition and level of moderation in conversations in those communities. Some participants completely avoided large generic subreddits such as r/news and r/politics and instead participated in relatively smaller niche political subreddits such as r/tuesday which is a relatively small center-right subreddit whose participants are derided as RINOs (Republicans In Name Only) for not being Republican enough and r/moderatepolitics, a moderate-sized non-partisan discussion subreddit where they could have more nuanced conversations. \footnote{r/tuesday and r/moderatepolitics had about 12,000 and 50,000 subscribers respectively at the time of conducting the study.} Note that these niche subreddits are not homogeneous partisan groups. They were relatively much smaller, well moderated, and frequented by members who similarly value cross-partisan interactions. P05 explained why he customizes the subreddits he participates in:

\textit{The reason why I follow some of these particular subreddits is just because people seem a little bit more reasonable in how they respond. You know, we probably both know that Reddit is primarily liberal, just in general. And you kind of have to go to specific subreddits if you want to get say like, right-leaning information or commentary. But some of the subreddits like r/Conservative and, you know, now banned, r/TheDonald, they're just so far over there. And the quality of discussion, in my opinion, is very, very low.} - P05

Through experience, some participants claimed to understand how their comments will be received depending on the type of subreddit in which they participate. A few others explained that they participated in subreddits that only partially aligned with their views which allowed them to have disagreements knowing that there was also some common ground.  

\textit{The r/Neoliberal one is a good one for me, just because I do dissent somewhat from some of the things they believe, but I also have a lot of common ground. So it's a space where I can have a lot of discussions with people who write, you know, at least they have similar moral frameworks, similar sort of ideological frameworks, even if some of the actual practices diverge a little bit there.} - P06

Even within subreddits, a few users are selective about which threads to engage in. For example, P04 explained that he recently started engaging in the open talk discussion threads on r/tuesday, rather than topic-specific ones. He reasoned that most of the subreddit ``regulars'' hung out there and were able to have deeper conversations since these threads do not usually get upvoted enough to show up on people's homefeeds and attract widespread attention from casual users. Generally, participants attest that identifying the right space to participate in tremendously affects all aspects of the discussion.

\subsubsection{Choosing who to talk to}
\label{intuitive}
All participants said that they viewed only the text of a comment by a user to decide whether to engage with them. Many participants described having an intuitive sense of how the conversation was going to unfold based on their reading of a users' initial comments. Some said that they could understand the `personality' of the user by reading between the lines to make quick judgments about whether to talk to them.

\textit{I can usually tell from the first comment--and I assume other people could as well--about the tone of the discussion with this person... Most of the time, it's just like, I know, that's going to be a bad convo, and that this is going to be more reasonable... I think I think it's probably about 90 or 95\%  of the time, the first comment generally identifies how the conversation is going to go.} - P05

Many participants also explained that they try not to enter into discussions with users who use profanity or strong emphasis words such as `obviously' or `clearly' when making suppositions on a topic. However, some participants also recalled times when they deliberately chose to engage with users making such statements when they were in a combative mood.

\subsection{Strategies adopted during cross-partisan discussions}

\label{strats}
\subsubsection{Establishing common ground and posing questions}

Participants recognize the current contentious political climate and are extra careful in how they communicate in cross-partisan discussions. Most participants reported that they typically start by signaling common ground with the user, highlighting parts of the argument that they agree with politely and respectfully. Then, they detail aspects that they disagree with while remaining extremely deliberate about how they frame their critiques, often posing them as questions. Many noted that they sometimes rewrite their comments multiple times to ensure that their views are conveyed accurately but without offending the people to whom they are talking. For example, P13, a high school teacher, described how she communicates with those she disagrees with.

\textit{I find that when I'm super careful about how I engage somebody whose opinions I differ with, the more careful I am, the better the conversation goes... rule number one when you have a parent-teacher conference is ``this is somebody's kid, say something nice.'' So for online, it's what do you agree with? What did this person say that you wholeheartedly agree with? and start from there. And then after that though, don't attack, [instead] question...} - P13

Thus, by establishing common ground and approaching conversation partners without a heavy gavel, users aim to signal that they are \textbf{open-minded and reflective}.

\subsubsection{Giving compliments and validating the interlocutor}

Some participants, upon sensing that a conversation has turned for the worse, typically give one last shot at reviving the conversation by explicitly complimenting or validating the interlocutor, as a sign of good faith. P18 explained how they sometimes try to correct the course of the conversation.

\textit{I think if someone's aggressive, but you can kind of sense that they do have the ability to have a better conversation, I think just being nice, I think not playing into their tricks, even validating them in a certain way helps... I love to say well, I agree with that. But I also have these things that I believe in and this and that, so I think actually validating them a little bit... you kind of show them your cooperation, they might actually come out to be more cooperative and I've seen that happen.} - P18

Thus, by acknowledging and validating the interlocutor, participants signal \textbf{good faith} to the interlocutors.

\subsubsection{Dispassionate argumentation}
 Many participants explicitly aimed to keep a calm and dispassionate demeanor when interacting with cross-partisans. Knowing that disagreements tend to engender strong emotions, participants keep the conversation on topic by focusing on the facts, providing arguments, and trying not to react emotionally to the interlocutor's arguments. For example, P04 described a particularly difficult conversation with a right-leaning user who argued that reports of police brutality in the US were overblown:
 
 \textit{I always try to start off with a very dispassionate response. And try to back up my claims with as much fact as possible and try to keep feelings out of it as much as possible, leave my worldview out of it as much as possible because we clearly don't share the same worldview. So I'm never going to be able to win that person over with that aspect, but just try to make it dispassionate.} - P04
 
 However, he conceded that being dispassionate on topics like the Black Lives Matter protests is especially difficult and instead, chose to disengage altogether.
In other instances, once participants sense that a user is becoming emotional in their replies, they swiftly disengage or concede the argument before it (potentially) devolves. For example, P12 said:
 
 \textit{There'll be times when I just stopped a conversation because someone's emotional. And I'll just concede the argument. There's no point in pushing somebody into a character attack when they're just getting emotionally invested in the argument.} - P12

Thus, users aim to remain dispassionate in their conversations to \textbf{maintain their own mental health and to prevent the conversation from potentially devolving into name-calling and character attacks}.

\subsubsection{Avoiding looking at the user's profile unless the conversation goes stale}
\label{no_profile}
Many users actively avoid learning more about their conversation partners by refraining from viewing their profile details such as karma points or past comments unless the conversation goes awry. By ignoring other possibly disturbing details about their conversation partner, participants focus their attention squarely on the argument that the person presents, not biased by their past opinions on other topics.

\textit{Normally, if I know that the other person that I talked to is a huge racist, or a sexist, or has some very, you know, skewed perspective on the world, then I would immediately want to stop talking to them... so I tend not to read the other person profile. I just try to, you know, discuss the topic with them, just that topic and nothing else... I don't want to know about that person, other than the things that are relevant for that discussion.} - P02

Others who do view past comments express doubts about whether knowing more about the user helps or hurts the conversation. For example, P04 was concerned about whether knowing a user's position outside the topic might prejudice him in the conversation.

\textit{I have [viewed past comments] in the past, especially with a name that I don't recognize, just to get an idea of what I'm getting myself into. But at the same time, I almost feel like that kind of almost prejudices me. And I almost want to have a narrowly scoped discussion that doesn't have the baggage of previous discussions or previous outside-of-this-subreddit's discussions.} - P04

Others, like P14 felt like knowing more about a user makes having a conversation with them difficult since the distance between their worldviews becomes more apparent. Looking at user profiles, most participants viewed karma points not as a predictive indicator of whether the user would be a good person to talk to but instead more as an explanatory variable when a conversation goes awry to make sense of the user's behavior.

Thus, by not viewing the user comment history or karma points, participants essentially try not to learn more about the user, ensuring that they discuss in \textbf{good faith and with an open mind without making assumptions}.

\subsubsection{Filtering the crowd}
Many participants recalled instances where their reply notifications ``blow up'' when multiple users angrily replied to their comments. In those instances, participants typically put on ``social blinders'' and focus on replying to only a specific individual.

\textit{Another thing I might do if I had a bad argument, but I liked one of the other people in the audience, and then everyone else is a bit [much]... I might sort of put like, social blinders on, and just tag that one person over and over again, to make it clear I'm only talking to them. Or I might continue the conversation in the chat box with them, Reddit has chat now, and continue in the DMs.} - P10

Unrestrained by the topic or other users in the subreddit, P16 found that DMs allow for users to switch topics and be more open. She noted that ``it just seems to me though, that direct message allows for a level of intimacy and being real.'' In other instances however, multiple participants reported that they have been directly targeted or harassed by others through DMs.

\subsubsection{Disengaging from the conversation}
By far, the most common reaction to a conversation that regresses into a personal attack or becomes combative is to disengage and exit the conversation.

\textit{I don't have to sit there and have somebody be ugly to me. That's not what I'm on the Internet for. I'm on the Internet to have fun and to be educated and not to be harassed.} - P11

Some participants use more stringent methods to disassociate themselves from the conversation by deleting their comments, reporting to the moderators, blocking the user and in rare cases, unsubscribing from the subreddit. It is important to note that disengaging is not a last resort action that participants take, oftentimes, disengaging is the first action that they take. Thus, to \textbf{safeguard their own mental health and to shield themselves from personal attacks}, participants simply disengage and walk away from the conversation.

\subsubsection{Counter strategies}
Not all strategies employed by the participants are conciliatory or aim to further the discussions. In some instances, participants said that they would counter by using aggressive or condescending language.

\textit{if I'm feeling petty, it's not like the right thing to do, but I like call them out in kind of a condescending way, I don't like using insults and things like that. But if I do want to be petty, it'll be more like, yeah, condescending or rhetorical questions, lighter, but still, I know, I shouldn't be talking like that.} - P08

In other situations, recognizing that the user they are talking to is angry, some users try to make them angrier.

\textit{Normally, when they're really mad and they go out of their way to like, target me. I normally just like, take the piss, you know, I kind of try to make them more mad... I don't confirm their prejudice. I just go, you know, oh man, look at this guy...haha... It's kind of stuff like that.} - P07

Others described using some of the tactics described by Jhaver and colleagues \cite{jhaver2018online} such as identity deception and sockpuppeting to counter hostility.

\textbf{Do these strategies work?} \textit{Sometimes.} Most participants acknowledged that while they do employ many of these strategies, the most effective approach in dealing with volatile conversations is to leave. Many recalled instances where they've tried to course correct a conversation only to make it worse. For example, P13 said:

\textit{I tried to engage once with somebody that vehement, and they just were, they just attacked. It was like, you know, it was like getting a text that's like, all caps from your mom. And it's just, you know, who needs that? So I'll just drop it. I don't reply. I just let it go.} - P13

Most participants explained that it is best to find another conversation to participate if their current conversation became worse. As P6 put it, ``when you invest a lot of energy into what is effectively an online discussion, it can sometimes feel like shoveling money into a fire.''

\subsection{Party Stereotyping and Dehumanization: Folk theories on what affects their conversation}
While we did not specifically ask participants why they thought their strategies did not always bear fruit, many participants had unprompted explanations of their own, specifically attributing party stereotyping and dehumanization as a cause for concern in cross-partisan discussions. \textcolor{black}{It was revealing (and frankly surprising) how well these folk theories matched with our cross-categorization and decategorization attempts through design.}

\subsubsection{Party Stereotyping}
Some participants attributed certain conversations going awry to stereotyping along party lines. In their experience, some users were quick to judge them as extreme liberal or conservative and project on them, what they perceive to be the typical characteristics of the group. P07 explained one such instance:

\textit{I think the worst one is where like, they kind of view you as the representation of like the right-wing or something. I'm not very conservative, but it's annoying when people are like, oh, you religious conservatives. Like, I'm not very religious. I'm not very conservative, they assume that like, you represent the whole like, you know, straw man of the entire wing} - P07

In other cases, participants were concerned about how a copartisan user supporting a position held by the participant may speak up for them. However, in doing so, they may provide reasons that are incongruent with the participant's own reasoning.

\textit{You end up with the problem of sometimes someone will say something as if he's speaking for you. But really, it's like, No, no, don't put me in there. [copartisan would say] ``And that's the issue..Republicans, Black people. And I'm sure everyone else here [agrees with me]'', please no no noooo! We are not the same, though.} - P06

Therefore, party stereotyping erases differences between individual group members (both ingroup and outgroup), leading users to \textbf{make broad assumptions of each other and affecting the ability to build common ground}.

\subsubsection{Dehumanization}
Contrasting with face-to-face interactions or interactions with people they personally know on other social media, many participants \textbf{attribute personal attacks} to dehumanizing effects afforded by anonymity on online platforms like Reddit.

\textit{Especially the anonymity that Reddit has, it's very easy for you to forget that that's a real person on the other side or for other people to forget that's a real person on the other side, you just start like throwing vitriol and people are just like, non-caring, like, will use any type of language to try and get their point across. And it's like, hey, I'm a human being, let's be at the very least cordial, we don't have to agree, but we should probably not try to like kill each other with words.} - P11

Either through personal experiences or subreddit rules or by reading the `Reddiquette' \footnote{informal norms that users are urged to subscribe to, \url{https://www.reddithelp.com/hc/en-us/articles/205926439}, ``Remember the human. When you communicate online, all you see is a computer screen. When talking to someone you might want to ask yourself "Would I say it to the person's face?" or "Would I get jumped if I said this to a buddy?"'}  which urges users to `remember the human', many participants recognize the need to view other users as human beings instead of a username on a screen.

\textit{I'm somebody who grew up with the internet evolving. I didn't start it when I was a kid. You know, I didn't have a phone in my hands until I was in my 20s. So I still go into every online conversation the way I would a real conversation. I'm constantly remembering that there's somebody on the other end. I'm consciously like this. I really pay attention to the words on the screen.} - P13

While participants understood the importance of remembering the human, they found it difficult to practice it online without other visual or auditory cues. Most participants felt that knowing more about the user and their interests would help view them as more complete human beings rather than just as someone who has strong political opinions. For example, P06 explained that the users he talks to online are strangers and that knowing more about them would humanize them:

\textit{It would be cool to know what kinds of stuff the other person's into, and just to maybe not put a face to it, but maybe, you know, at least see some additional humanity behind what is otherwise a username and text.} - P06

However, many of the participants who acknowledged the importance of `seeing the human' remained deeply skeptical of knowing too much about their conversation partner for fear that extra information may distract or bias the conversation. For example, later, when asked if he would like to know more about users he talks to, P06 said:

\textit{In a sense, I don't want to know very much about the person other than that they are a good partner or conversationalist or whatever... I wouldn't want to know anything about the person, their race, I wouldn't want to know their gender, I wouldn't want to know shit... I think beyond including resources that clue people into someone being a good debate partner, the other information becomes more so distracting or brings about expectations that will guide the conversation in a way that is not based on the substance of the argument itself.} - P06

Thus, many participants appear to navigate the following paradox: knowing too little, you risk dehumanizing them. Knowing too much, you risk the integrity of the conversation---and usually, participants lean toward minimizing the additional information they know about the user.

\subsection{How do users consider the extra information provided by the designs?}

\subsubsection{Shared subreddits}

\paragraph{\underline{Potential for humanizing users}}
Many participants stated, often enthusiastically, that viewing shared subreddits on the user card would remind them that there was a real person, a human being, on the other side of a conversation. For P08, shared subreddits would make them feel more connected to the user who is otherwise just a random stranger, and would likely reduce anger and negative emotions.

\textit{I think this would be very humanizing. I think you can see what kind of, you know, interests they have on Reddit outside of politics and the conversation that you're having... if it's happening in a negative or a politically charged conversation... you know when you're talking with someone anonymous, you can be a lot ruder, a lot more condescending, and there's not really consequences to it, but when you see this, I think for me, it would reduce my anger or my negative emotions.} - P08

\paragraph{\underline{Potential for reducing stereotyping}}
Other participants explained that highlighting commonalities could help bridge the gap and see the person in a (relatively) more positive light.

\textit{I think [shared subreddits] is helpful because outside of the political spectrum people do have common interests. So my feeling might be, well, okay, maybe this person is not so bad. They like technology, they share the same interests in sports.} - P17

\paragraph{\underline{Potential for fostering good faith and common ground}}
Many participants felt that viewing the subreddits they shared with another user would help establish for themselves some common ground with the user. They explained that in conversations that get particularly heated, knowing that they share a common interest would help to build some goodwill.  

\textit{I think that could sponsor a little more goodwill among people, like, even if you have two people that are vehemently arguing with each other and calling each other--you know, flipping each other off verbally--if they find out that, oh, you have an ATV too, or a four-wheeler and you'd like to go out, it could sponsor a little bit more goodwill, which I think could ultimately lead to better conversations, for sure. And it's a good idea.} - P05

\paragraph{\underline{Concerns}}
Some participants were concerned that there might be few instances where the users share common subreddits, however, our data analysis (in Section \ref{feas}) revealed that a sizable number of cross-partisan and copartisan pairs participate in at least one common non-political subreddit. Also, a few participants explained that they would be inclined to look at the user card only if it was someone that they recognize or have spoken to earlier. They thought that this information would be less useful for one-time interactions. 

\subsubsection{Active subreddits}
\paragraph{\underline{Potential for reducing stereotyping}}
As expected, some participants explained that knowing the other subreddits in which their interlocutor participates would help reassure them that the person is not fixated on politics and has other interests as well.

\textit{[Showing active subreddits would help because] that'll tell me if you're not stuck in a particular way, that they do have other interests that could influence their thought process.} - P17

A few participants liked that they could quickly get an idea of the kinds of subreddits in which the user participates. They explained that currently, they needed to scroll through a reverse-chronological list of their past comments to get a sense of where they participated.

\paragraph{\underline{Concerns}}
Participants expressed two significant concerns with the active subreddits component. First, some participants felt that some active subreddits could present them in a negative light. They feared that when others view that they participated in a fun subreddit such as a meme subreddit, they may not take their political arguments seriously or worse still, use their participation in those subreddits to discredit their arguments.

\textit{[Would not like active subreddits because] I don't want it to be like, Oh, this person's trying to describe to me economics and they browse like, I don't know, but just the Jojo subreddit all day. You know, it's a very easy path for like judgment, I guess.} - P07

The other concern was that, in its current form, the active subreddits component simply displayed too much information about their activity on the site. Many participants suggested that providing a way to customize the subreddits shown on their card or to opt-out of displaying the active subreddits will allay these concerns.

When you're first reading a comment from people it's like an interaction at a bar--you want to give them enough so that they come over, but you don't want to sell them the house. - P13

However, another factor that might compound these concerns is that this design when deployed as a browser extension would result in an information asymmetry, users may not even know that their interlocutor is using the extension to view information about them. This is a serious concern that we expand on in Section \ref{asymmetry}.

\subsubsection{Comment highlights}
Some participants liked the idea of viewing different dimensions of a person based on their top comments in other subreddits. Some even recounted past comments that became viral or were gilded (awarded Reddit gold) by other users which they would be proud to highlight on the user card. However, many participants raised two major concerns. First, participants were concerned that comment highlights based on karma points could produce a biased view of the person and cause easy judgment. They explained that most top-voted comments were either extremely opinionated, controversial, or partisan which might provide fodder for more conflict.

\textit{If someone is passionate about one thing and not passionate about the other, or somebody could have an extreme opinion about one thing and not about another. So you would see the most extreme thing, maybe, as their top comment, and then now you get to judge people on their most extreme opinion. I don't think that would be a very good idea.} - P07

Others noted the comments would likely distract them from the actual conversation or may contain outdated beliefs which may color the viewer's opinion about them. Another major concern was that participants expressed feeling self-conscious about the information revealed in the comment highlights. It is telling that all four participants stating this concern were either female or genderqueer (P8, P13, P14, P18). They felt that the comment highlights made their comments more public and their profile more open to scrutiny. P14 expressed that they would likely choose how they word their comments carefully because of the increased visibility. P18 explained that she liked the way past comments were structured on the Reddit profile page, a simple list of past comments in reverse chronological order. It afforded her privacy by not being very organized or accessible.
 
 \textit{[I like the profile page] because I know that it's not necessarily that open and always accessible and when people are touching on touchy topics and they are expressing themselves, [they] might want to keep some sort of an anonymity, I feel like having things more presented in a way that shares more information could actually be a problem.} - P18

Further, P18 expressed concern that her views on one topic may be used against her when she is discussing other issues and said that she would likely have to make throwaway accounts to prevent users from connecting issues. Similarly, while P13 did not express specific concerns about the design, she had earlier described a prior experience where users racially abused her after finding out that she was a Black woman from a photo she had previously uploaded. This component likely exacerbates these concerns by increasing the visibility of specific comments.

\subsubsection{Karma points and awards}
\paragraph{\underline{Potential as a basic/weak good faith indicator}}
Many interview participants expressed that it did not matter if their interlocutor amassed high karma points and awards, as they used it not so much to determine if the person posted quality comments, but to simply indicate if an interlocutor was a troll.

\paragraph{\underline{Concerns}}
Many participants explained that high karma points only indicated that the person makes good jokes or puns and it said nothing about the quality of someone's views. This somewhat lukewarm response to karma points is in line with Massarani's observation that while Redditors place some value on karma scores, they are also suspicious of users with very high scores \cite{massanari2015participatory}. Almost all right-leaning participants pointed out that since Reddit has more liberal users, the karma points and awards usually only reveal how liberal the user is and therefore might bias the conversation. Hearing this initial feedback, we converted our karma indicator to display only if they had less than 100 karma for use as a very basic good faith indicator. Later participants told us that this information, coupled with information on the age of the account, was enough to identify troll behavior.

\section{Discussion}

\subsection{Education vs entertainment in cross-partisan discussions}
We find that participants engage in political discussions both to educate and to entertain. Depending on external factors such as time constraints and outside news cycles, the same user may engage in relatively serious political discussions or may casually peruse the site and join in casual banter, satire, and trolling. This finding is also in line with past work that shows that trolling behavior is context-dependent and not an immutable individual characteristic \cite{cheng2017anyone}.

The two goals may be at odds with each other. The same comments may be appreciated differently by people seeking education vs. entertainment. Interventions that aim to coach users to talk to the other side (such as \cite{yeomans2020conversational}) might help produce comments that are more effective for education than entertainment. Participants may also be more receptive to such coaching when they are primarily motivated by education rather than entertainment. On the other hand, the two goals can also be complementary. None of our participants joined Reddit for its political content and most had significant interests in other non-political subreddit topics. By hosting something for everybody, Reddit likely allows casual political observers who happen to peruse the site for other reasons to engage in political talk. Also, many participants commented that they often switched to lighter content when conversations go awry or when they simply needed a break from a heavy discussion. We speculate that this easy access to fun and entertaining content has therapeutic effects and serves to lighten the after-effects of serious political discussions. This recuperative function is particularly important given the participants' concern about the emotional toll of these conversations. 

\subsection{Unintended consequences of cross-partisan discourse?}

The outcomes of cross-partisan interactions, both online and offline, have been typically evaluated in terms of highly valued outcomes such as political participation and political efficacy. However, little is known about the effects of cross-partisan interactions on users' emotional and mental well-being. From our interviews, we observe that most participants were wary about the discussions' repercussions on their mental health and employed multiple strategies to negate these effects. Thus, we call on researchers to attend to the psychological effects of participating in these discussions in addition to studying normative democratic outcomes, especially in these highly polarized times. 

We observe that many participants, as a form of mental self-preservation, aim to have dispassionate discussions and sometimes even preemptively disengage if they feel that they or their interlocutors are getting emotional, for fear that emotions could devolve into name-calling. They make a distinction between being ``emotional'' and ``rational''. However, this hyper-rational, impersonal style of deliberation could have unintended consequences. Firstly, research suggests that taking emotions out of political discussions does not necessarily lead to more rational outcomes; while anger spurs aversion and leads to close-mindedness, when the emotional response is anxiety, people seek new perspectives and become open to compromise \cite{mackuen2010civic}. Anger, on the other hand, also increases political participation \cite{valentino2011election}. Secondly, as Young notes, ``a norm of dispassionateness dismisses and devalues embodied forms of expression, emotion, and figurative expressions. People's contributions to a discussion tend to be excluded from serious consideration not because of what is said, but how it is said.'' \cite{young2002inclusion} Clearly, this limits whose views are engaged with in cross-partisan interactions; users who are directly affected by the discussed issues likely passionately voice their opinions while those that are unaffected likely remain detached. Thus, the views of users who have the highest stakes may be less attended to. Finally, pure reasoning, with its emphasis on rationality as opposed to passion is known to be also exclusionary towards members of disadvantaged groups and individuals with less formal education as this form of communication is deliberately learned and developed \cite{mouffe2000politics}. Important questions around the outcomes of these conversations remain; does this kind of cross-partisan discourse contribute positively to building a deliberative democracy? In its current form, the prevailing hostility, toll on mental health, and the possible unintended consequences of participants' strategies to have deliberative discussions suggest otherwise--at least on Reddit.

\subsection{Impacts of information about interlocutors: humanizing, stereotyping, judging, and attacking}

In our interviews before showing the design probes, participants readily acknowledged that knowing more about others could be humanizing, allowing them to ``see a little more humanity in what is otherwise a username and text'' (P06). However, in current practice, participants described that they exclusively focused on the comment text and not on the author of the comment due to the fear that they may become prejudiced by viewing the author's past behavior or positions. Given that the Reddit interface does not provide the means to only see humanizing information while avoiding prejudice-inducing information, participants resolve this issue by simply not viewing user profiles entirely. This reduces opportunities to build common ground and trust. We aimed to address this issue by showing potentially humanizing information about the user through our designs.

Upon viewing the user card, as expected, participants indicated that it would alter how they participate in conversations, making them consider both the comment and the comment's author when responding. In the case of shared subreddits, many predicted that this shift could humanize the interlocutors and promote goodwill, with participants expressing that they would be more mindful of their behavior and more charitable of their interlocutors' potential transgressions. However, participants also expressed many significant concerns about other ways that the information might be used. With active subreddits, they worried that other users might judge/stereotype them negatively for participating in casual subreddits such as meme subreddits and also felt this component disclosed too much information about them. For comment highlights, female and minority participants were especially concerned about how these comments could provide more fodder to attack them. Some of these concerns can be addressed by providing users with more control over what information is shown about them on the user card.

However, more broadly, focusing attention on the user profiles, while potentially humanizing especially when users share group memberships, could have major negative implications especially for female and minority users by increasing visibility and inviting increased scrutiny on their profiles. Although participants expressed frustration over Reddit's anonymity providing a safe harbor for trolls, they also appreciated how this anonymity allowed them to express opinions without being targeted.

\subsection{Challenges and opportunities for future designs to improve cross-partisan discourse}
Given the concerning feedback that we received, we decided not to proceed with building the proposed browser extension. Instead, we detail challenges in designing to improve cross-partisan discourse and opportunities we see to move forward in this space.

\subsubsection{Countering the different forms of hostility in cross-partisan discussions}

Partisan attacks and name-calling during cross-partisan discussions are commonplace online. Our designs aimed to minimize such occurrences by highlighting non-political group memberships to offset the effects of outgroup categorizations. However, it is important to consider other forms of hostility. For example, determined users can search through interlocutors' activity history to find material to disrupt the discussion and attack them. These concerns are particularly significant for many of our female and minority participants for whom partisan hostility often interacts with sexism and racism in cross-partisan interactions. The culture of harassment based on toxic conceptions of race, gender and sexual identities supported by Reddit's design and governance, which Massanari terms as ``toxic technocultures''  \cite{massanari2017gamergate}, negatively influence and exacerbate the hostility already prevalent in these political discussions. Future work on designs to improve cross-partisan discourse should attend to the multiple forms of hostility prevalent and how the designs to reduce hostility may differentially impact members of disadvantaged and marginalized social groups.

\subsubsection{Countering party-stereotypes without revealing user information}
Our designs aimed to cross-categorize and decategorize at an individual level by making certain user activity more visible. However, as we showed in Section \ref{feas}, not all user interacting pairs have common group memberships. Further, in the previous section, we highlighted some concerns about revealing user information, and one user went so far as to say she would make throwaway accounts to disrupt that. Thus, alternate approaches to de-stereotype without calling attention to individual profiles may be more effective. For example, Alher et al. \cite{ahler2018parties} found that people consistently overestimate the extent to which party supporters belong to party-stereotypical groups, sometimes by over 300\% (for example, atheists for Democrats and evangelical for Republicans) and correcting these misconceptions led to significant reductions in outparty hostility. Similarly, we could surface subreddit memberships in aggregate to counter some of these extreme stereotypes. For example, by showing that only (a surprisingly low) 5.34\% of Reddit users who participate in r/Conservative also participate in r/Christianity (based on 2019 Reddit comment data).

\subsubsection{Intervening in cross-partisan discussions}
In recent years, researchers have developed algorithms to detect when a conversation is likely to go awry to encourage either the moderators or the conversation participants to possibly take course correction measures \cite{chang2019trouble,liu2018forecasting}. However, from our interviews, participants' own attempts at de-escalating often either have little/no effect or cause more harm (Section \ref{strats}). \textcolor{black}{Designers aiming to intervene in individual discussions must evaluate if and when to intervene, taking into account the possible adverse effects of their interventions. We recommend that such systems allow users to choose if they want intervention in the first place so that they help facilitate and prolong discussions that the users actually want to participate in and help exit ones that they do not.} However, given the low rates of success for turning around a conversation and the possibility of unintended harm, we urge designers to explore preventive measures such as improving community norms around deliberation rather than corrective measures to improve individual discussions.

\subsubsection{Information Asymmetry}\label{asymmetry}
Our designs aimed to provide more information about interlocutors to facilitate better discussions. As outside researchers do not have access to the site, these designs are typically deployed as browser extensions or external apps. However, such a deployment would result in some users (who download the extension) having easy access to information about others, while other users may not even know that their interlocutors have access to their information. Further, even if the extension allowed users to customize or remove content on their user cards, users first would need to know that such an extension exists and download it. This will likely compound concerns that users already have about revealing more information about them. Given that cross-partisan interactions often turn into adversarial situations, one approach could be to apply an affirmative consent lens to design, centering individual agency with interactions structured around consent that is voluntary, informed, revertible, specific, and unburdensome \cite{im2021yes}. Thus, a possible modification could be that users be able to view subreddit participation details of only other extension users who consent to information sharing. This change may necessitate a user recruitment strategy where extension users have a high likelihood of interacting with each other. To maximize the chances of such interaction, the deployment could be targeted to users participating in a particular subreddit. 

\section{Limitations}

Our study focuses on cross-partisan discussions on Reddit only; future work on other platforms will surely improve our understanding. Given our participants' strong support for showing shared group memberships between Reddit users who are essentially strangers, we expect that showing such connections on Facebook, especially between weak ties, will have a similar impact. However, we expect that showing other active group memberships will have little impact on Facebook as users already have access to some individuating information about others in the form of a real name, profile picture and cover photo, unlike on Reddit where users typically only identify themselves using a username.

Our study is US-centric and was conducted in highly polarized times, during the lead-up to one of the most contentious US presidential elections in history. In less polarized countries/settings, partisan identities will be less salient; we speculate that these designs would have smaller effects on reducing hostility in interactions, as partisan group dynamics is unlikely to be the cause for the hostility. Alternately, approaches aimed at establishing more deliberative discussion norms through example setting \cite{sukumaran2011normative} may be more effective as these norms might face less resistance from the hostile partisan norms that we observe today.

While in this work, we have focused on cross-partisan political discussions online, we do not contend that cross-partisan interactions are more important or should take primacy over other forms of political discourse that in some cases specifically exclude dissenting voices. As, Mansbridge et. al  \cite{Mansbridge_Bohman_Chambers_Christiano_Fung_Parkinson_Thompson_Warren_2012} note:

\textit{Activist interactions in social movement enclaves are often highly partisan, closed to opposing ideas, and disrespectful of opponents. Yet the intensity of interaction and even the exclusion of opposing ideas in such enclaves create the fertile, protected hothouses sometimes necessary to generate counter-hegemonic ideas. These ideas then may play powerful roles in the broader deliberative system, substantively improving an eventual democratic decision.}

\section{Conclusion}
In this work, we have explored how users navigate the contentious political climate in the US to engage in cross-partisan discussions. We find that participants have different, multiple motivations for engaging in these interactions; sometimes they prefer serious deliberative discussions and other times, they look for entertainment and banter. These different motivations coupled with the hyper-partisan environment present challenges to participants seeking to engage with ``the other side''. Through experience, participants have developed multiple strategies to foster good conversations. From our design probes, we observe that participants find shared non-political subreddit memberships of their interlocutors humanizing, however, sharing other details such as other group subreddit memberships and past top comments raise significant concerns around privacy and misuse. 

\section{Acknowledgements}
We thank Libby Hemphill, Angela Schöpke-Gonzalez and Vaishnav Kameswaran for their detailed reviews and suggestions on earlier drafts of the paper. We also thank the anonymous reviewers for their incredibly useful feedback over multiple revisions which has shaped this work immensely. Ashwin Rajadesingan is supported by a Facebook Fellowship. This material is based upon work supported by the National Science Foundation under Grant No. IIS-1717688.  

\bibliographystyle{ACM-Reference-Format}
\bibliography{sample-base}


\begin{thebibliography}{83}


\ifx \showCODEN    \undefined \def \showCODEN     #1{\unskip}     \fi
\ifx \showDOI      \undefined \def \showDOI       #1{#1}\fi
\ifx \showISBNx    \undefined \def \showISBNx     #1{\unskip}     \fi
\ifx \showISBNxiii \undefined \def \showISBNxiii  #1{\unskip}     \fi
\ifx \showISSN     \undefined \def \showISSN      #1{\unskip}     \fi
\ifx \showLCCN     \undefined \def \showLCCN      #1{\unskip}     \fi
\ifx \shownote     \undefined \def \shownote      #1{#1}          \fi
\ifx \showarticletitle \undefined \def \showarticletitle #1{#1}   \fi
\ifx \showURL      \undefined \def \showURL       {\relax}        \fi
\providecommand\bibfield[2]{#2}
\providecommand\bibinfo[2]{#2}
\providecommand\natexlab[1]{#1}
\providecommand\showeprint[2][]{arXiv:#2}

\bibitem[\protect\citeauthoryear{Ahler and Sood}{Ahler and Sood}{2018}]%
        {ahler2018parties}
\bibfield{author}{\bibinfo{person}{Douglas~J Ahler} {and}
  \bibinfo{person}{Gaurav Sood}.} \bibinfo{year}{2018}\natexlab{}.
\newblock \showarticletitle{The parties in our heads: Misperceptions about
  party composition and their consequences}.
\newblock \bibinfo{journal}{\emph{The Journal of Politics}}
  \bibinfo{volume}{80}, \bibinfo{number}{3} (\bibinfo{year}{2018}),
  \bibinfo{pages}{964--981}.
\newblock


\bibitem[\protect\citeauthoryear{Allison and Bussey}{Allison and
  Bussey}{2020}]%
        {allison2020communal}
\bibfield{author}{\bibinfo{person}{Kimberley Allison} {and}
  \bibinfo{person}{Kay Bussey}.} \bibinfo{year}{2020}\natexlab{}.
\newblock \showarticletitle{Communal quirks and circlejerks: A taxonomy of
  processes contributing to insularity in online communities}. In
  \bibinfo{booktitle}{\emph{Proceedings of the International AAAI Conference on
  Web and Social Media}}.
\newblock


\bibitem[\protect\citeauthoryear{Anderson and Auxier}{Anderson and
  Auxier}{2020}]%
        {Anderson_2020}
\bibfield{author}{\bibinfo{person}{Monica Anderson} {and}
  \bibinfo{person}{Brooke Auxier}.} \bibinfo{year}{2020}\natexlab{}.
\newblock \showarticletitle{55\% of U.S. social media users say they are
  ‘worn out’ by political posts and discussions}.
\newblock \bibinfo{journal}{\emph{Pew Research Center
  https://www.pewresearch.org/fact-tank/2020/08/19/55-of-u-s-social-media-users-say-they-are-worn-out-by-political-posts-and-discussions/}}
  (\bibinfo{year}{2020}).
\newblock


\bibitem[\protect\citeauthoryear{B{\"a}chtiger, Dryzek, Mansbridge, and
  Warren}{B{\"a}chtiger et~al\mbox{.}}{2018}]%
        {bachtiger2018deliberative}
\bibfield{author}{\bibinfo{person}{Andr{\'e} B{\"a}chtiger},
  \bibinfo{person}{John~S Dryzek}, \bibinfo{person}{Jane Mansbridge}, {and}
  \bibinfo{person}{M Warren}.} \bibinfo{year}{2018}\natexlab{}.
\newblock \showarticletitle{Deliberative Democracy}.
\newblock \bibinfo{journal}{\emph{The Oxford handbook of deliberative
  democracy}} (\bibinfo{year}{2018}), \bibinfo{pages}{1}.
\newblock


\bibitem[\protect\citeauthoryear{Baumgartner, Zannettou, Keegan, Squire, and
  Blackburn}{Baumgartner et~al\mbox{.}}{2020}]%
        {baumgartner2020pushshift}
\bibfield{author}{\bibinfo{person}{Jason Baumgartner}, \bibinfo{person}{Savvas
  Zannettou}, \bibinfo{person}{Brian Keegan}, \bibinfo{person}{Megan Squire},
  {and} \bibinfo{person}{Jeremy Blackburn}.} \bibinfo{year}{2020}\natexlab{}.
\newblock \showarticletitle{The pushshift reddit dataset}. In
  \bibinfo{booktitle}{\emph{Proceedings of the International AAAI Conference on
  Web and Social Media}}, Vol.~\bibinfo{volume}{14}. \bibinfo{pages}{830--839}.
\newblock


\bibitem[\protect\citeauthoryear{Brewer}{Brewer}{1984}]%
        {brewer1984beyond}
\bibfield{author}{\bibinfo{person}{Marilynn~B Brewer}.}
  \bibinfo{year}{1984}\natexlab{}.
\newblock \showarticletitle{Beyond the contact hypothesis: Theoretical
  perspectives on desegregation}.
\newblock \bibinfo{journal}{\emph{Groups in contact: The psychology of
  desegregation}} (\bibinfo{year}{1984}), \bibinfo{pages}{281--302}.
\newblock


\bibitem[\protect\citeauthoryear{Brewer}{Brewer}{2000}]%
        {brewer2000reducing}
\bibfield{author}{\bibinfo{person}{Marilynn~B Brewer}.}
  \bibinfo{year}{2000}\natexlab{}.
\newblock \showarticletitle{Reducing Prejudice Through Cross-Categorization:
  Effects}.
\newblock \bibinfo{journal}{\emph{Reducing prejudice and discrimination}}
  (\bibinfo{year}{2000}), \bibinfo{pages}{165--185}.
\newblock


\bibitem[\protect\citeauthoryear{Chandrasekharan, Pavalanathan, Srinivasan,
  Glynn, Eisenstein, and Gilbert}{Chandrasekharan et~al\mbox{.}}{2017}]%
        {chandrasekharan2017you}
\bibfield{author}{\bibinfo{person}{Eshwar Chandrasekharan},
  \bibinfo{person}{Umashanthi Pavalanathan}, \bibinfo{person}{Anirudh
  Srinivasan}, \bibinfo{person}{Adam Glynn}, \bibinfo{person}{Jacob
  Eisenstein}, {and} \bibinfo{person}{Eric Gilbert}.}
  \bibinfo{year}{2017}\natexlab{}.
\newblock \showarticletitle{You can't stay here: The efficacy of reddit's 2015
  ban examined through hate speech}.
\newblock \bibinfo{journal}{\emph{Proceedings of the ACM on Human-Computer
  Interaction}} \bibinfo{volume}{1}, \bibinfo{number}{CSCW}
  (\bibinfo{year}{2017}), \bibinfo{pages}{1--22}.
\newblock


\bibitem[\protect\citeauthoryear{Chandrasekharan, Samory, Jhaver, Charvat,
  Bruckman, Lampe, Eisenstein, and Gilbert}{Chandrasekharan
  et~al\mbox{.}}{2018}]%
        {chandrasekharan2018internet}
\bibfield{author}{\bibinfo{person}{Eshwar Chandrasekharan},
  \bibinfo{person}{Mattia Samory}, \bibinfo{person}{Shagun Jhaver},
  \bibinfo{person}{Hunter Charvat}, \bibinfo{person}{Amy Bruckman},
  \bibinfo{person}{Cliff Lampe}, \bibinfo{person}{Jacob Eisenstein}, {and}
  \bibinfo{person}{Eric Gilbert}.} \bibinfo{year}{2018}\natexlab{}.
\newblock \showarticletitle{The Internet's hidden rules: An empirical study of
  Reddit norm violations at micro, meso, and macro scales}.
\newblock \bibinfo{journal}{\emph{Proceedings of the ACM on Human-Computer
  Interaction}} \bibinfo{volume}{2}, \bibinfo{number}{CSCW}
  (\bibinfo{year}{2018}), \bibinfo{pages}{1--25}.
\newblock


\bibitem[\protect\citeauthoryear{Chang and Danescu-Niculescu-Mizil}{Chang and
  Danescu-Niculescu-Mizil}{2019}]%
        {chang2019trouble}
\bibfield{author}{\bibinfo{person}{Jonathan~P Chang} {and}
  \bibinfo{person}{Cristian Danescu-Niculescu-Mizil}.}
  \bibinfo{year}{2019}\natexlab{}.
\newblock \showarticletitle{Trouble on the Horizon: Forecasting the Derailment
  of Online Conversations as they Develop}. In
  \bibinfo{booktitle}{\emph{Proceedings of the 2019 Conference on Empirical
  Methods in Natural Language Processing and the 9th International Joint
  Conference on Natural Language Processing (EMNLP-IJCNLP)}}.
  \bibinfo{pages}{4745--4756}.
\newblock


\bibitem[\protect\citeauthoryear{Charmaz}{Charmaz}{2006}]%
        {charmaz2006constructing}
\bibfield{author}{\bibinfo{person}{Kathy Charmaz}.}
  \bibinfo{year}{2006}\natexlab{}.
\newblock \bibinfo{booktitle}{\emph{Constructing grounded theory: A practical
  guide through qualitative analysis}}.
\newblock \bibinfo{publisher}{sage}.
\newblock


\bibitem[\protect\citeauthoryear{Cheng, Bernstein, Danescu-Niculescu-Mizil, and
  Leskovec}{Cheng et~al\mbox{.}}{2017}]%
        {cheng2017anyone}
\bibfield{author}{\bibinfo{person}{Justin Cheng}, \bibinfo{person}{Michael
  Bernstein}, \bibinfo{person}{Cristian Danescu-Niculescu-Mizil}, {and}
  \bibinfo{person}{Jure Leskovec}.} \bibinfo{year}{2017}\natexlab{}.
\newblock \showarticletitle{Anyone can become a troll: Causes of trolling
  behavior in online discussions}. In \bibinfo{booktitle}{\emph{Proceedings of
  the 2017 ACM conference on computer supported cooperative work and social
  computing}}. \bibinfo{pages}{1217--1230}.
\newblock


\bibitem[\protect\citeauthoryear{Crisp and Hewstone}{Crisp and
  Hewstone}{2007}]%
        {crisp2007multiple}
\bibfield{author}{\bibinfo{person}{Richard~J Crisp} {and}
  \bibinfo{person}{Miles Hewstone}.} \bibinfo{year}{2007}\natexlab{}.
\newblock \showarticletitle{Multiple social categorization}.
\newblock \bibinfo{journal}{\emph{Advances in experimental social psychology}}
  \bibinfo{volume}{39} (\bibinfo{year}{2007}), \bibinfo{pages}{163--254}.
\newblock


\bibitem[\protect\citeauthoryear{Crisp, Walsh, and Hewstone}{Crisp
  et~al\mbox{.}}{2006}]%
        {crisp2006crossed}
\bibfield{author}{\bibinfo{person}{Richard~J Crisp}, \bibinfo{person}{Judi
  Walsh}, {and} \bibinfo{person}{Miles Hewstone}.}
  \bibinfo{year}{2006}\natexlab{}.
\newblock \showarticletitle{Crossed categorization in common ingroup contexts}.
\newblock \bibinfo{journal}{\emph{Personality and Social Psychology Bulletin}}
  \bibinfo{volume}{32}, \bibinfo{number}{9} (\bibinfo{year}{2006}),
  \bibinfo{pages}{1204--1218}.
\newblock


\bibitem[\protect\citeauthoryear{De~Choudhury and De}{De~Choudhury and
  De}{2014}]%
        {de2014mental}
\bibfield{author}{\bibinfo{person}{Munmun De~Choudhury} {and}
  \bibinfo{person}{Sushovan De}.} \bibinfo{year}{2014}\natexlab{}.
\newblock \showarticletitle{Mental health discourse on reddit: Self-disclosure,
  social support, and anonymity}. In \bibinfo{booktitle}{\emph{Proceedings of
  the International AAAI Conference on Web and Social Media}},
  Vol.~\bibinfo{volume}{8}.
\newblock


\bibitem[\protect\citeauthoryear{DellaPosta, Shi, and Macy}{DellaPosta
  et~al\mbox{.}}{2015}]%
        {dellaposta2015liberals}
\bibfield{author}{\bibinfo{person}{Daniel DellaPosta}, \bibinfo{person}{Yongren
  Shi}, {and} \bibinfo{person}{Michael Macy}.} \bibinfo{year}{2015}\natexlab{}.
\newblock \showarticletitle{Why do liberals drink lattes?}
\newblock \bibinfo{journal}{\emph{Amer. J. Sociology}} (\bibinfo{year}{2015}).
\newblock


\bibitem[\protect\citeauthoryear{Druckman, Klar, Krupnikov, Levendusky, and
  Ryan}{Druckman et~al\mbox{.}}{2021}]%
        {druckman_misestimating}
\bibfield{author}{\bibinfo{person}{James~N. Druckman}, \bibinfo{person}{Samara
  Klar}, \bibinfo{person}{Yanna Krupnikov}, \bibinfo{person}{Matthew
  Levendusky}, {and} \bibinfo{person}{John~Barry Ryan}.}
  \bibinfo{year}{2021}\natexlab{}.
\newblock \showarticletitle{(Mis-)Estimating Affective Polarization}.
\newblock \bibinfo{journal}{\emph{The Journal of Politics}}
  \bibinfo{volume}{0}, \bibinfo{number}{ja} (\bibinfo{year}{2021}),
  \bibinfo{pages}{null}.
\newblock
\urldef\tempurl%
\url{https://doi.org/10.1086/715603}
\showDOI{\tempurl}
\showeprint{https://doi.org/10.1086/715603}


\bibitem[\protect\citeauthoryear{Duggan and Smith}{Duggan and Smith}{2016}]%
        {Duggan_Smith_2016}
\bibfield{author}{\bibinfo{person}{M. Duggan} {and} \bibinfo{person}{A.
  Smith}.} \bibinfo{year}{2016}\natexlab{}.
\newblock \showarticletitle{The tone of social media discussions around
  politics}.
\newblock \bibinfo{journal}{\emph{Pew Research Center http://www. pewinternet.
  org/2016/10/25/the-tone-of-social-media-discussions-around-politics}}
  (\bibinfo{year}{2016}).
\newblock


\bibitem[\protect\citeauthoryear{Faridani, Bitton, Ryokai, and
  Goldberg}{Faridani et~al\mbox{.}}{2010}]%
        {faridani2010opinion}
\bibfield{author}{\bibinfo{person}{Siamak Faridani}, \bibinfo{person}{Ephrat
  Bitton}, \bibinfo{person}{Kimiko Ryokai}, {and} \bibinfo{person}{Ken
  Goldberg}.} \bibinfo{year}{2010}\natexlab{}.
\newblock \showarticletitle{Opinion space: a scalable tool for browsing online
  comments}. In \bibinfo{booktitle}{\emph{Proceedings of the SIGCHI Conference
  on Human Factors in Computing Systems}}. \bibinfo{pages}{1175--1184}.
\newblock


\bibitem[\protect\citeauthoryear{Fiesler, Jiang, McCann, Frye, and
  Brubaker}{Fiesler et~al\mbox{.}}{2018}]%
        {fiesler2018reddit}
\bibfield{author}{\bibinfo{person}{Casey Fiesler}, \bibinfo{person}{Jialun
  Jiang}, \bibinfo{person}{Joshua McCann}, \bibinfo{person}{Kyle Frye}, {and}
  \bibinfo{person}{Jed Brubaker}.} \bibinfo{year}{2018}\natexlab{}.
\newblock \showarticletitle{Reddit rules! characterizing an ecosystem of
  governance}. In \bibinfo{booktitle}{\emph{Proceedings of the International
  AAAI Conference on Web and Social Media}}, Vol.~\bibinfo{volume}{12}.
\newblock


\bibitem[\protect\citeauthoryear{Gaver, Dunne, and Pacenti}{Gaver
  et~al\mbox{.}}{1999}]%
        {gaver1999design}
\bibfield{author}{\bibinfo{person}{Bill Gaver}, \bibinfo{person}{Tony Dunne},
  {and} \bibinfo{person}{Elena Pacenti}.} \bibinfo{year}{1999}\natexlab{}.
\newblock \showarticletitle{Design: cultural probes}.
\newblock \bibinfo{journal}{\emph{interactions}} \bibinfo{volume}{6},
  \bibinfo{number}{1} (\bibinfo{year}{1999}), \bibinfo{pages}{21--29}.
\newblock


\bibitem[\protect\citeauthoryear{Gervais}{Gervais}{2015}]%
        {gervais2015incivility}
\bibfield{author}{\bibinfo{person}{Bryan~T Gervais}.}
  \bibinfo{year}{2015}\natexlab{}.
\newblock \showarticletitle{Incivility online: Affective and behavioral
  reactions to uncivil political posts in a web-based experiment}.
\newblock \bibinfo{journal}{\emph{Journal of Information Technology \&
  Politics}} \bibinfo{volume}{12}, \bibinfo{number}{2} (\bibinfo{year}{2015}),
  \bibinfo{pages}{167--185}.
\newblock


\bibitem[\protect\citeauthoryear{Grevet, Terveen, and Gilbert}{Grevet
  et~al\mbox{.}}{2014}]%
        {grevet2014managing}
\bibfield{author}{\bibinfo{person}{Catherine Grevet}, \bibinfo{person}{Loren~G
  Terveen}, {and} \bibinfo{person}{Eric Gilbert}.}
  \bibinfo{year}{2014}\natexlab{}.
\newblock \showarticletitle{Managing political differences in social media}. In
  \bibinfo{booktitle}{\emph{Proceedings of the 17th ACM conference on Computer
  supported cooperative work \& social computing}}.
  \bibinfo{pages}{1400--1408}.
\newblock


\bibitem[\protect\citeauthoryear{Habermas and Habermas}{Habermas and
  Habermas}{1991}]%
        {habermas1991structural}
\bibfield{author}{\bibinfo{person}{Jurgen Habermas} {and}
  \bibinfo{person}{J{\"u}rgen Habermas}.} \bibinfo{year}{1991}\natexlab{}.
\newblock \bibinfo{booktitle}{\emph{The structural transformation of the public
  sphere: An inquiry into a category of bourgeois society}}.
\newblock \bibinfo{publisher}{MIT press}.
\newblock


\bibitem[\protect\citeauthoryear{Hancock, Toma, and Fenner}{Hancock
  et~al\mbox{.}}{2008}]%
        {hancock2008know}
\bibfield{author}{\bibinfo{person}{Jeffrey~T Hancock},
  \bibinfo{person}{Catalina~L Toma}, {and} \bibinfo{person}{Kate Fenner}.}
  \bibinfo{year}{2008}\natexlab{}.
\newblock \showarticletitle{I know something you don't: the use of asymmetric
  personal information for interpersonal advantage}. In
  \bibinfo{booktitle}{\emph{Proceedings of the 2008 ACM conference on Computer
  supported cooperative work}}. \bibinfo{pages}{413--416}.
\newblock


\bibitem[\protect\citeauthoryear{Hendriks, Dryzek, and Hunold}{Hendriks
  et~al\mbox{.}}{2007}]%
        {hendriks2007turning}
\bibfield{author}{\bibinfo{person}{Carolyn~M Hendriks}, \bibinfo{person}{John~S
  Dryzek}, {and} \bibinfo{person}{Christian Hunold}.}
  \bibinfo{year}{2007}\natexlab{}.
\newblock \showarticletitle{Turning up the heat: Partisanship in deliberative
  innovation}.
\newblock \bibinfo{journal}{\emph{Political studies}} \bibinfo{volume}{55},
  \bibinfo{number}{2} (\bibinfo{year}{2007}), \bibinfo{pages}{362--383}.
\newblock


\bibitem[\protect\citeauthoryear{Hogg and Turner}{Hogg and Turner}{1987}]%
        {hogg1987intergroup}
\bibfield{author}{\bibinfo{person}{Michael~A Hogg} {and}
  \bibinfo{person}{John~C Turner}.} \bibinfo{year}{1987}\natexlab{}.
\newblock \showarticletitle{Intergroup behaviour, self-stereotyping and the
  salience of social categories}.
\newblock \bibinfo{journal}{\emph{British Journal of Social Psychology}}
  \bibinfo{volume}{26}, \bibinfo{number}{4} (\bibinfo{year}{1987}),
  \bibinfo{pages}{325--340}.
\newblock


\bibitem[\protect\citeauthoryear{Huddy and Bankert}{Huddy and Bankert}{2017}]%
        {huddy2017political}
\bibfield{author}{\bibinfo{person}{Leonie Huddy} {and} \bibinfo{person}{Alexa
  Bankert}.} \bibinfo{year}{2017}\natexlab{}.
\newblock \showarticletitle{Political partisanship as a social identity}.
\newblock In \bibinfo{booktitle}{\emph{Oxford research encyclopedia of
  politics}}.
\newblock


\bibitem[\protect\citeauthoryear{Im, Dimond, Berton, Lee, Mustelier, Ackerman,
  and Gilbert}{Im et~al\mbox{.}}{2021}]%
        {im2021yes}
\bibfield{author}{\bibinfo{person}{Jane Im}, \bibinfo{person}{Jill Dimond},
  \bibinfo{person}{Melody Berton}, \bibinfo{person}{Una Lee},
  \bibinfo{person}{Katherine Mustelier}, \bibinfo{person}{Mark Ackerman}, {and}
  \bibinfo{person}{Eric Gilbert}.} \bibinfo{year}{2021}\natexlab{}.
\newblock \showarticletitle{Yes: Affirmative Consent as a Theoretical Framework
  for Understanding and Imagining Social Platforms}.
\newblock  (\bibinfo{year}{2021}).
\newblock


\bibitem[\protect\citeauthoryear{Iyengar, Lelkes, Levendusky, Malhotra, and
  Westwood}{Iyengar et~al\mbox{.}}{2019}]%
        {iyengar2019origins}
\bibfield{author}{\bibinfo{person}{Shanto Iyengar}, \bibinfo{person}{Yphtach
  Lelkes}, \bibinfo{person}{Matthew Levendusky}, \bibinfo{person}{Neil
  Malhotra}, {and} \bibinfo{person}{Sean~J Westwood}.}
  \bibinfo{year}{2019}\natexlab{}.
\newblock \showarticletitle{The origins and consequences of affective
  polarization in the United States}.
\newblock \bibinfo{journal}{\emph{Annual Review of Political Science}}
  \bibinfo{volume}{22} (\bibinfo{year}{2019}), \bibinfo{pages}{129--146}.
\newblock


\bibitem[\protect\citeauthoryear{Iyengar, Sood, and Lelkes}{Iyengar
  et~al\mbox{.}}{2012}]%
        {iyengar2012affect}
\bibfield{author}{\bibinfo{person}{Shanto Iyengar}, \bibinfo{person}{Gaurav
  Sood}, {and} \bibinfo{person}{Yphtach Lelkes}.}
  \bibinfo{year}{2012}\natexlab{}.
\newblock \showarticletitle{Affect, not ideologya social identity perspective
  on polarization}.
\newblock \bibinfo{journal}{\emph{Public opinion quarterly}}
  \bibinfo{volume}{76}, \bibinfo{number}{3} (\bibinfo{year}{2012}),
  \bibinfo{pages}{405--431}.
\newblock


\bibitem[\protect\citeauthoryear{Jhaver, Appling, Gilbert, and Bruckman}{Jhaver
  et~al\mbox{.}}{2019}]%
        {jhaver2019did}
\bibfield{author}{\bibinfo{person}{Shagun Jhaver},
  \bibinfo{person}{Darren~Scott Appling}, \bibinfo{person}{Eric Gilbert}, {and}
  \bibinfo{person}{Amy Bruckman}.} \bibinfo{year}{2019}\natexlab{}.
\newblock \showarticletitle{" Did You Suspect the Post Would be Removed?"
  Understanding User Reactions to Content Removals on Reddit}.
\newblock \bibinfo{journal}{\emph{Proceedings of the ACM on human-computer
  interaction}} \bibinfo{volume}{3}, \bibinfo{number}{CSCW}
  (\bibinfo{year}{2019}), \bibinfo{pages}{1--33}.
\newblock


\bibitem[\protect\citeauthoryear{Jhaver, Ghoshal, Bruckman, and Gilbert}{Jhaver
  et~al\mbox{.}}{2018}]%
        {jhaver2018online}
\bibfield{author}{\bibinfo{person}{Shagun Jhaver}, \bibinfo{person}{Sucheta
  Ghoshal}, \bibinfo{person}{Amy Bruckman}, {and} \bibinfo{person}{Eric
  Gilbert}.} \bibinfo{year}{2018}\natexlab{}.
\newblock \showarticletitle{Online harassment and content moderation: The case
  of blocklists}.
\newblock \bibinfo{journal}{\emph{ACM Transactions on Computer-Human
  Interaction (TOCHI)}} \bibinfo{volume}{25}, \bibinfo{number}{2}
  (\bibinfo{year}{2018}), \bibinfo{pages}{1--33}.
\newblock


\bibitem[\protect\citeauthoryear{Kim and Kim}{Kim and Kim}{2008}]%
        {Kim_Kim_2008}
\bibfield{author}{\bibinfo{person}{Joohan Kim} {and} \bibinfo{person}{Eun~Joo
  Kim}.} \bibinfo{year}{2008}\natexlab{}.
\newblock \showarticletitle{Theorizing dialogic deliberation: Everyday
  political talk as communicative action and dialogue}.
\newblock \bibinfo{journal}{\emph{Communication Theory}} \bibinfo{volume}{18},
  \bibinfo{number}{1} (\bibinfo{year}{2008}), \bibinfo{pages}{51–70}.
\newblock


\bibitem[\protect\citeauthoryear{Klar, Krupnikov, and Ryan}{Klar
  et~al\mbox{.}}{2018}]%
        {klar2018affective}
\bibfield{author}{\bibinfo{person}{Samara Klar}, \bibinfo{person}{Yanna
  Krupnikov}, {and} \bibinfo{person}{John~Barry Ryan}.}
  \bibinfo{year}{2018}\natexlab{}.
\newblock \showarticletitle{Affective polarization or partisan disdain?
  Untangling a dislike for the opposing party from a dislike of partisanship}.
\newblock \bibinfo{journal}{\emph{Public Opinion Quarterly}}
  \bibinfo{volume}{82}, \bibinfo{number}{2} (\bibinfo{year}{2018}),
  \bibinfo{pages}{379--390}.
\newblock


\bibitem[\protect\citeauthoryear{Kraut, Levine, Escobar, and
  Herda{\u{g}}delen}{Kraut et~al\mbox{.}}{2020}]%
        {kraut2020makes}
\bibfield{author}{\bibinfo{person}{Robert~E Kraut}, \bibinfo{person}{John~M
  Levine}, \bibinfo{person}{Marisol~Martinez Escobar}, {and}
  \bibinfo{person}{Ama{\c{c}} Herda{\u{g}}delen}.}
  \bibinfo{year}{2020}\natexlab{}.
\newblock \showarticletitle{What Makes People Feel Close to Online Groups? The
  Roles of Group Attributes and Group Types}. In
  \bibinfo{booktitle}{\emph{Proceedings of the International AAAI Conference on
  Web and Social Media}}, Vol.~\bibinfo{volume}{14}. \bibinfo{pages}{382--392}.
\newblock


\bibitem[\protect\citeauthoryear{Kriplean, Morgan, Freelon, Borning, and
  Bennett}{Kriplean et~al\mbox{.}}{2012a}]%
        {Kriplean_Morgan_Freelon_Borning_Bennett_2012}
\bibfield{author}{\bibinfo{person}{Travis Kriplean}, \bibinfo{person}{Jonathan
  Morgan}, \bibinfo{person}{Deen Freelon}, \bibinfo{person}{Alan Borning},
  {and} \bibinfo{person}{Lance Bennett}.} \bibinfo{year}{2012}\natexlab{a}.
\newblock \showarticletitle{Supporting reflective public thought with
  considerit}. In \bibinfo{booktitle}{\emph{Proceedings of the ACM 2012
  conference on Computer Supported Cooperative Work}}.
  \bibinfo{publisher}{ACM}.
\newblock


\bibitem[\protect\citeauthoryear{Kriplean, Toomim, Morgan, Borning, and
  Ko}{Kriplean et~al\mbox{.}}{2012b}]%
        {Kriplean_Toomim_Morgan_Borning_Ko_2012}
\bibfield{author}{\bibinfo{person}{Travis Kriplean}, \bibinfo{person}{Michael
  Toomim}, \bibinfo{person}{Jonathan Morgan}, \bibinfo{person}{Alan Borning},
  {and} \bibinfo{person}{Andrew Ko}.} \bibinfo{year}{2012}\natexlab{b}.
\newblock \showarticletitle{Is this what you meant?: promoting listening on the
  web with reflect}. In \bibinfo{booktitle}{\emph{Proceedings of the SIGCHI
  Conference on Human Factors in Computing Systems}}. \bibinfo{publisher}{ACM},
  \bibinfo{pages}{1559–1568}.
\newblock


\bibitem[\protect\citeauthoryear{Leavitt}{Leavitt}{2015}]%
        {leavitt2015throwaway}
\bibfield{author}{\bibinfo{person}{Alex Leavitt}.}
  \bibinfo{year}{2015}\natexlab{}.
\newblock \showarticletitle{" This is a Throwaway Account" Temporary Technical
  Identities and Perceptions of Anonymity in a Massive Online Community}. In
  \bibinfo{booktitle}{\emph{Proceedings of the 18th ACM conference on computer
  supported cooperative work \& social computing}}. \bibinfo{pages}{317--327}.
\newblock


\bibitem[\protect\citeauthoryear{Levendusky}{Levendusky}{2020}]%
        {levendusky2020our}
\bibfield{author}{\bibinfo{person}{Matthew~S Levendusky}.}
  \bibinfo{year}{2020}\natexlab{}.
\newblock \showarticletitle{Our common bonds: Using what Americans share to
  help bridge the partisan divide}.
\newblock \bibinfo{journal}{\emph{Unpublished Manuscript, University of
  Pennsylvania}} (\bibinfo{year}{2020}).
\newblock


\bibitem[\protect\citeauthoryear{Liu, Guberman, Hemphill, and Culotta}{Liu
  et~al\mbox{.}}{2018}]%
        {liu2018forecasting}
\bibfield{author}{\bibinfo{person}{P Liu}, \bibinfo{person}{J Guberman},
  \bibinfo{person}{L Hemphill}, {and} \bibinfo{person}{A Culotta}.}
  \bibinfo{year}{2018}\natexlab{}.
\newblock \showarticletitle{Forecasting the presence and intensity of hostility
  on Instagram using linguistic and social features}. In
  \bibinfo{booktitle}{\emph{Proceedings of the 12th International Conference on
  Web and Social Media}}.
\newblock


\bibitem[\protect\citeauthoryear{MacKuen, Wolak, Keele, and Marcus}{MacKuen
  et~al\mbox{.}}{2010}]%
        {mackuen2010civic}
\bibfield{author}{\bibinfo{person}{Michael MacKuen}, \bibinfo{person}{Jennifer
  Wolak}, \bibinfo{person}{Luke Keele}, {and} \bibinfo{person}{George~E
  Marcus}.} \bibinfo{year}{2010}\natexlab{}.
\newblock \showarticletitle{Civic engagements: Resolute partisanship or
  reflective deliberation}.
\newblock \bibinfo{journal}{\emph{American Journal of Political Science}}
  \bibinfo{volume}{54}, \bibinfo{number}{2} (\bibinfo{year}{2010}),
  \bibinfo{pages}{440--458}.
\newblock


\bibitem[\protect\citeauthoryear{Mansbridge, Bohman, Chambers, Christiano,
  Fung, Parkinson, Thompson, and Warren}{Mansbridge et~al\mbox{.}}{2012}]%
  {Mansbridge_Bohman_Chambers_Christiano_Fung_Parkinson_Thompson_Warren_2012}
\bibfield{author}{\bibinfo{person}{Jane Mansbridge}, \bibinfo{person}{James
  Bohman}, \bibinfo{person}{Simone Chambers}, \bibinfo{person}{Thomas
  Christiano}, \bibinfo{person}{Archon Fung}, \bibinfo{person}{John Parkinson},
  \bibinfo{person}{Dennis~F. Thompson}, {and} \bibinfo{person}{Mark~E.
  Warren}.} \bibinfo{year}{2012}\natexlab{}.
\newblock \showarticletitle{A systemic approach to deliberative democracy}.
\newblock \bibinfo{journal}{\emph{Deliberative systems: Deliberative democracy
  at the large scale}} (\bibinfo{year}{2012}), \bibinfo{pages}{1–26}.
\newblock


\bibitem[\protect\citeauthoryear{Marwick and Boyd}{Marwick and Boyd}{2011}]%
        {marwick2011tweet}
\bibfield{author}{\bibinfo{person}{Alice~E Marwick} {and}
  \bibinfo{person}{Danah Boyd}.} \bibinfo{year}{2011}\natexlab{}.
\newblock \showarticletitle{I tweet honestly, I tweet passionately: Twitter
  users, context collapse, and the imagined audience}.
\newblock \bibinfo{journal}{\emph{New media \& society}} \bibinfo{volume}{13},
  \bibinfo{number}{1} (\bibinfo{year}{2011}), \bibinfo{pages}{114--133}.
\newblock


\bibitem[\protect\citeauthoryear{Mason}{Mason}{2016}]%
        {mason2016cross}
\bibfield{author}{\bibinfo{person}{Lilliana Mason}.}
  \bibinfo{year}{2016}\natexlab{}.
\newblock \showarticletitle{A cross-cutting calm: How social sorting drives
  affective polarization}.
\newblock \bibinfo{journal}{\emph{Public Opinion Quarterly}}
  \bibinfo{volume}{80}, \bibinfo{number}{S1} (\bibinfo{year}{2016}),
  \bibinfo{pages}{351--377}.
\newblock


\bibitem[\protect\citeauthoryear{Massanari}{Massanari}{2017}]%
        {massanari2017gamergate}
\bibfield{author}{\bibinfo{person}{Adrienne Massanari}.}
  \bibinfo{year}{2017}\natexlab{}.
\newblock \showarticletitle{\# Gamergate and The Fappening: How Reddit’s
  algorithm, governance, and culture support toxic technocultures}.
\newblock \bibinfo{journal}{\emph{New Media \& Society}} \bibinfo{volume}{19},
  \bibinfo{number}{3} (\bibinfo{year}{2017}), \bibinfo{pages}{329--346}.
\newblock


\bibitem[\protect\citeauthoryear{Massanari}{Massanari}{2015}]%
        {massanari2015participatory}
\bibfield{author}{\bibinfo{person}{Adrienne~Lynne Massanari}.}
  \bibinfo{year}{2015}\natexlab{}.
\newblock \showarticletitle{Participatory culture, community, and play}.
\newblock \bibinfo{journal}{\emph{Learning from}} (\bibinfo{year}{2015}).
\newblock


\bibitem[\protect\citeauthoryear{Mouffe}{Mouffe}{2000}]%
        {mouffe2000politics}
\bibfield{author}{\bibinfo{person}{Chantal Mouffe}.}
  \bibinfo{year}{2000}\natexlab{}.
\newblock \showarticletitle{Politics and Passions}.
\newblock \bibinfo{journal}{\emph{Ethical Perspectives}} \bibinfo{volume}{7},
  \bibinfo{number}{2-3} (\bibinfo{year}{2000}), \bibinfo{pages}{146--150}.
\newblock


\bibitem[\protect\citeauthoryear{Muddiman and Stroud}{Muddiman and
  Stroud}{2017}]%
        {muddiman2017news}
\bibfield{author}{\bibinfo{person}{Ashley Muddiman} {and}
  \bibinfo{person}{Natalie~Jomini Stroud}.} \bibinfo{year}{2017}\natexlab{}.
\newblock \showarticletitle{News values, cognitive biases, and partisan
  incivility in comment sections}.
\newblock \bibinfo{journal}{\emph{Journal of communication}}
  \bibinfo{volume}{67}, \bibinfo{number}{4} (\bibinfo{year}{2017}),
  \bibinfo{pages}{586--609}.
\newblock


\bibitem[\protect\citeauthoryear{Munson, Lee, and Resnick}{Munson
  et~al\mbox{.}}{2013}]%
        {munson2013encouraging}
\bibfield{author}{\bibinfo{person}{Sean~A Munson}, \bibinfo{person}{Stephanie~Y
  Lee}, {and} \bibinfo{person}{Paul Resnick}.} \bibinfo{year}{2013}\natexlab{}.
\newblock \showarticletitle{Encouraging reading of diverse political viewpoints
  with a browser widget}. In \bibinfo{booktitle}{\emph{Seventh international
  aaai conference on weblogs and social media}}.
\newblock


\bibitem[\protect\citeauthoryear{Munson and Resnick}{Munson and
  Resnick}{2010}]%
        {munson2010presenting}
\bibfield{author}{\bibinfo{person}{Sean~A Munson} {and} \bibinfo{person}{Paul
  Resnick}.} \bibinfo{year}{2010}\natexlab{}.
\newblock \showarticletitle{Presenting diverse political opinions: how and how
  much}. In \bibinfo{booktitle}{\emph{Proceedings of the SIGCHI conference on
  human factors in computing systems}}. \bibinfo{pages}{1457--1466}.
\newblock


\bibitem[\protect\citeauthoryear{Nelimarkka, Rancy, Grygiel, and
  Semaan}{Nelimarkka et~al\mbox{.}}{2019a}]%
        {nelimarkka2019re}
\bibfield{author}{\bibinfo{person}{Matti Nelimarkka},
  \bibinfo{person}{Jean~Philippe Rancy}, \bibinfo{person}{Jennifer Grygiel},
  {and} \bibinfo{person}{Bryan Semaan}.} \bibinfo{year}{2019}\natexlab{a}.
\newblock \showarticletitle{(Re) Design to Mitigate Political Polarization:
  Reflecting Habermas' ideal communication space in the United States of
  America and Finland}.
\newblock \bibinfo{journal}{\emph{Proceedings of the ACM on Human-Computer
  Interaction}} \bibinfo{volume}{3}, \bibinfo{number}{CSCW}
  (\bibinfo{year}{2019}), \bibinfo{pages}{1--25}.
\newblock


\bibitem[\protect\citeauthoryear{Nelimarkka, Rancy, Grygiel, and
  Semaan}{Nelimarkka et~al\mbox{.}}{2019b}]%
        {Nelimarkka_Rancy_Grygiel_Semaan_2019}
\bibfield{author}{\bibinfo{person}{Matti Nelimarkka},
  \bibinfo{person}{Jean~Philippe Rancy}, \bibinfo{person}{Jennifer Grygiel},
  {and} \bibinfo{person}{Bryan Semaan}.} \bibinfo{year}{2019}\natexlab{b}.
\newblock \showarticletitle{(Re) Design to Mitigate Political Polarization:
  Reflecting Habermas’ ideal communication space in the United States of
  America and Finland}.
\newblock \bibinfo{journal}{\emph{Proceedings of the ACM on Human-Computer
  Interaction}} \bibinfo{volume}{3}, \bibinfo{number}{CSCW}
  (\bibinfo{year}{2019}), \bibinfo{pages}{141}.
\newblock


\bibitem[\protect\citeauthoryear{Pariser}{Pariser}{2011}]%
        {pariser2011filter}
\bibfield{author}{\bibinfo{person}{Eli Pariser}.}
  \bibinfo{year}{2011}\natexlab{}.
\newblock \bibinfo{booktitle}{\emph{The filter bubble: What the Internet is
  hiding from you}}.
\newblock \bibinfo{publisher}{Penguin UK}.
\newblock


\bibitem[\protect\citeauthoryear{Pettigrew}{Pettigrew}{1998}]%
        {pettigrew1998intergroup}
\bibfield{author}{\bibinfo{person}{Thomas~F Pettigrew}.}
  \bibinfo{year}{1998}\natexlab{}.
\newblock \showarticletitle{Intergroup contact theory}.
\newblock \bibinfo{journal}{\emph{Annual review of psychology}}
  \bibinfo{volume}{49}, \bibinfo{number}{1} (\bibinfo{year}{1998}),
  \bibinfo{pages}{65--85}.
\newblock


\bibitem[\protect\citeauthoryear{Postmes and Baym}{Postmes and Baym}{2005}]%
        {postmes2005intergroup}
\bibfield{author}{\bibinfo{person}{Tom Postmes} {and} \bibinfo{person}{Nancy
  Baym}.} \bibinfo{year}{2005}\natexlab{}.
\newblock \showarticletitle{Intergroup dimensions of the Internet}.
\newblock \bibinfo{journal}{\emph{Intergroup communication: Multiple
  perspectives}}  \bibinfo{volume}{2} (\bibinfo{year}{2005}),
  \bibinfo{pages}{213--240}.
\newblock


\bibitem[\protect\citeauthoryear{Postmes, Spears, and Lea}{Postmes
  et~al\mbox{.}}{2002}]%
        {postmes2002intergroup}
\bibfield{author}{\bibinfo{person}{Tom Postmes}, \bibinfo{person}{Russell
  Spears}, {and} \bibinfo{person}{Martin Lea}.}
  \bibinfo{year}{2002}\natexlab{}.
\newblock \showarticletitle{Intergroup differentiation in computer-mediated
  communication: Effects of depersonalization.}
\newblock \bibinfo{journal}{\emph{Group Dynamics: Theory, Research, and
  Practice}} \bibinfo{volume}{6}, \bibinfo{number}{1} (\bibinfo{year}{2002}),
  \bibinfo{pages}{3}.
\newblock


\bibitem[\protect\citeauthoryear{Rains, Kenski, Coe, and Harwood}{Rains
  et~al\mbox{.}}{2017}]%
        {rains2017incivility}
\bibfield{author}{\bibinfo{person}{Stephen~A Rains}, \bibinfo{person}{Kate
  Kenski}, \bibinfo{person}{Kevin Coe}, {and} \bibinfo{person}{Jake Harwood}.}
  \bibinfo{year}{2017}\natexlab{}.
\newblock \showarticletitle{Incivility and political identity on the Internet:
  Intergroup factors as predictors of incivility in discussions of news
  online}.
\newblock \bibinfo{journal}{\emph{Journal of Computer-Mediated Communication}}
  \bibinfo{volume}{22}, \bibinfo{number}{4} (\bibinfo{year}{2017}),
  \bibinfo{pages}{163--178}.
\newblock


\bibitem[\protect\citeauthoryear{Rajadesingan, Budak, and Resnick}{Rajadesingan
  et~al\mbox{.}}{2021}]%
        {rajadesingan2021political}
\bibfield{author}{\bibinfo{person}{Ashwin Rajadesingan}, \bibinfo{person}{Ceren
  Budak}, {and} \bibinfo{person}{Paul Resnick}.}
  \bibinfo{year}{2021}\natexlab{}.
\newblock \showarticletitle{Political Discussion is Abundant in Non-political
  Subreddits (and Less Toxic)}. In \bibinfo{booktitle}{\emph{Proceedings of the
  International AAAI Conference on Web and Social Media}},
  Vol.~\bibinfo{volume}{15}. \bibinfo{pages}{525--536}.
\newblock


\bibitem[\protect\citeauthoryear{Rajadesingan, Resnick, and Budak}{Rajadesingan
  et~al\mbox{.}}{2020}]%
        {rajadesingan2020quick}
\bibfield{author}{\bibinfo{person}{Ashwin Rajadesingan}, \bibinfo{person}{Paul
  Resnick}, {and} \bibinfo{person}{Ceren Budak}.}
  \bibinfo{year}{2020}\natexlab{}.
\newblock \showarticletitle{Quick, Community-Specific Learning: How Distinctive
  Toxicity Norms Are Maintained in Political Subreddits}. In
  \bibinfo{booktitle}{\emph{Proceedings of the International AAAI Conference on
  Web and Social Media}}, Vol.~\bibinfo{volume}{14}. \bibinfo{pages}{557--568}.
\newblock


\bibitem[\protect\citeauthoryear{Reicher, Spears, and Postmes}{Reicher
  et~al\mbox{.}}{1995}]%
        {Reicher_Spears_Postmes_1995}
\bibfield{author}{\bibinfo{person}{Stephen~D. Reicher},
  \bibinfo{person}{Russell Spears}, {and} \bibinfo{person}{Tom Postmes}.}
  \bibinfo{year}{1995}\natexlab{}.
\newblock \showarticletitle{A social identity model of deindividuation
  phenomena}.
\newblock \bibinfo{journal}{\emph{European review of social psychology}}
  \bibinfo{volume}{6}, \bibinfo{number}{1} (\bibinfo{year}{1995}),
  \bibinfo{pages}{161–198}.
\newblock


\bibitem[\protect\citeauthoryear{Seering, Fang, Damasco, Chen, Sun, and
  Kaufman}{Seering et~al\mbox{.}}{2019}]%
        {Seering_Fang_Damasco_Chen_Sun_Kaufman_2019}
\bibfield{author}{\bibinfo{person}{Joseph Seering}, \bibinfo{person}{Tianmi
  Fang}, \bibinfo{person}{Luca Damasco}, \bibinfo{person}{Mianhong’Cherie’
  Chen}, \bibinfo{person}{Likang Sun}, {and} \bibinfo{person}{Geoff Kaufman}.}
  \bibinfo{year}{2019}\natexlab{}.
\newblock \showarticletitle{Designing User Interface Elements to Improve the
  Quality and Civility of Discourse in Online Commenting Behaviors}. In
  \bibinfo{booktitle}{\emph{Proceedings of the 2019 CHI Conference on Human
  Factors in Computing Systems}}. \bibinfo{publisher}{ACM},
  \bibinfo{pages}{606}.
\newblock


\bibitem[\protect\citeauthoryear{Seering, Kraut, and Dabbish}{Seering
  et~al\mbox{.}}{2017}]%
        {seering2017shaping}
\bibfield{author}{\bibinfo{person}{Joseph Seering}, \bibinfo{person}{Robert
  Kraut}, {and} \bibinfo{person}{Laura Dabbish}.}
  \bibinfo{year}{2017}\natexlab{}.
\newblock \showarticletitle{Shaping pro and anti-social behavior on twitch
  through moderation and example-setting}. In
  \bibinfo{booktitle}{\emph{Proceedings of the 2017 ACM conference on computer
  supported cooperative work and social computing}}. \bibinfo{pages}{111--125}.
\newblock


\bibitem[\protect\citeauthoryear{Semaan, Faucett, Robertson, Maruyama, and
  Douglas}{Semaan et~al\mbox{.}}{2015}]%
        {Semaan_Faucett_Robertson_Maruyama_Douglas_2015}
\bibfield{author}{\bibinfo{person}{Bryan Semaan}, \bibinfo{person}{Heather
  Faucett}, \bibinfo{person}{Scott~P. Robertson}, \bibinfo{person}{Misa
  Maruyama}, {and} \bibinfo{person}{Sara Douglas}.}
  \bibinfo{year}{2015}\natexlab{}.
\newblock \showarticletitle{Designing Political Deliberation Environments to
  Support Interactions in the Public Sphere} \emph{(\bibinfo{series}{CHI
  ’15})}. \bibinfo{publisher}{ACM}, \bibinfo{pages}{3167–3176}.
\newblock
\showISBNx{978-1-4503-3145-6}


\bibitem[\protect\citeauthoryear{Semaan, Robertson, Douglas, and
  Maruyama}{Semaan et~al\mbox{.}}{2014a}]%
        {semaan2014social}
\bibfield{author}{\bibinfo{person}{Bryan~C Semaan}, \bibinfo{person}{Scott~P
  Robertson}, \bibinfo{person}{Sara Douglas}, {and} \bibinfo{person}{Misa
  Maruyama}.} \bibinfo{year}{2014}\natexlab{a}.
\newblock \showarticletitle{Social media supporting political deliberation
  across multiple public spheres: towards depolarization}. In
  \bibinfo{booktitle}{\emph{Proceedings of the 17th ACM conference on Computer
  supported cooperative work \& social computing}}.
  \bibinfo{pages}{1409--1421}.
\newblock


\bibitem[\protect\citeauthoryear{Semaan, Robertson, Douglas, and
  Maruyama}{Semaan et~al\mbox{.}}{2014b}]%
        {Semaan_Robertson_Douglas_Maruyama_2014}
\bibfield{author}{\bibinfo{person}{Bryan~C. Semaan}, \bibinfo{person}{Scott~P.
  Robertson}, \bibinfo{person}{Sara Douglas}, {and} \bibinfo{person}{Misa
  Maruyama}.} \bibinfo{year}{2014}\natexlab{b}.
\newblock \showarticletitle{Social media supporting political deliberation
  across multiple public spheres: towards depolarization}. In
  \bibinfo{booktitle}{\emph{Proceedings of the 17th ACM conference on Computer
  supported cooperative work \& social computing}}. \bibinfo{publisher}{ACM},
  \bibinfo{pages}{1409–1421}.
\newblock


\bibitem[\protect\citeauthoryear{Settle}{Settle}{2018}]%
        {settle2018frenemies}
\bibfield{author}{\bibinfo{person}{Jaime~E Settle}.}
  \bibinfo{year}{2018}\natexlab{}.
\newblock \bibinfo{booktitle}{\emph{Frenemies: How social media polarizes
  America}}.
\newblock \bibinfo{publisher}{Cambridge University Press}.
\newblock


\bibitem[\protect\citeauthoryear{Shafranek}{Shafranek}{2019}]%
        {shafranek2019political}
\bibfield{author}{\bibinfo{person}{Richard~M Shafranek}.}
  \bibinfo{year}{2019}\natexlab{}.
\newblock \showarticletitle{Political considerations in nonpolitical decisions:
  a conjoint analysis of roommate choice}.
\newblock \bibinfo{journal}{\emph{Political Behavior}} (\bibinfo{year}{2019}),
  \bibinfo{pages}{1--30}.
\newblock


\bibitem[\protect\citeauthoryear{Shah, Cho, Eveland~Jr, and Kwak}{Shah
  et~al\mbox{.}}{2005}]%
        {shah2005information}
\bibfield{author}{\bibinfo{person}{Dhavan~V Shah}, \bibinfo{person}{Jaeho Cho},
  \bibinfo{person}{William~P Eveland~Jr}, {and} \bibinfo{person}{Nojin Kwak}.}
  \bibinfo{year}{2005}\natexlab{}.
\newblock \showarticletitle{Information and expression in a digital age:
  Modeling Internet effects on civic participation}.
\newblock \bibinfo{journal}{\emph{Communication research}}
  \bibinfo{volume}{32}, \bibinfo{number}{5} (\bibinfo{year}{2005}),
  \bibinfo{pages}{531--565}.
\newblock


\bibitem[\protect\citeauthoryear{Shipman and Marshall}{Shipman and
  Marshall}{1999}]%
        {shipman1999formality}
\bibfield{author}{\bibinfo{person}{Frank~M Shipman} {and}
  \bibinfo{person}{Catherine~C Marshall}.} \bibinfo{year}{1999}\natexlab{}.
\newblock \showarticletitle{Formality considered harmful: Experiences, emerging
  themes, and directions on the use of formal representations in interactive
  systems}.
\newblock \bibinfo{journal}{\emph{Computer Supported Cooperative Work (CSCW)}}
  \bibinfo{volume}{8}, \bibinfo{number}{4} (\bibinfo{year}{1999}),
  \bibinfo{pages}{333--352}.
\newblock


\bibitem[\protect\citeauthoryear{Shum et~al\mbox{.}}{Shum
  et~al\mbox{.}}{2008}]%
        {shum2008cohere}
\bibfield{author}{\bibinfo{person}{Simon~Buckingham Shum} {et~al\mbox{.}}}
  \bibinfo{year}{2008}\natexlab{}.
\newblock \showarticletitle{Cohere: Towards web 2.0 argumentation}.
\newblock \bibinfo{journal}{\emph{COMMA}}  \bibinfo{volume}{8}
  (\bibinfo{year}{2008}), \bibinfo{pages}{97--108}.
\newblock


\bibitem[\protect\citeauthoryear{Suhay, Bello-Pardo, and Maurer}{Suhay
  et~al\mbox{.}}{2018}]%
        {suhay2018polarizing}
\bibfield{author}{\bibinfo{person}{Elizabeth Suhay}, \bibinfo{person}{Emily
  Bello-Pardo}, {and} \bibinfo{person}{Brianna Maurer}.}
  \bibinfo{year}{2018}\natexlab{}.
\newblock \showarticletitle{The polarizing effects of online partisan
  criticism: Evidence from two experiments}.
\newblock \bibinfo{journal}{\emph{The International Journal of Press/Politics}}
  \bibinfo{volume}{23}, \bibinfo{number}{1} (\bibinfo{year}{2018}),
  \bibinfo{pages}{95--115}.
\newblock


\bibitem[\protect\citeauthoryear{Sukumaran, Vezich, McHugh, and Nass}{Sukumaran
  et~al\mbox{.}}{2011}]%
        {sukumaran2011normative}
\bibfield{author}{\bibinfo{person}{Abhay Sukumaran}, \bibinfo{person}{Stephanie
  Vezich}, \bibinfo{person}{Melanie McHugh}, {and} \bibinfo{person}{Clifford
  Nass}.} \bibinfo{year}{2011}\natexlab{}.
\newblock \showarticletitle{Normative influences on thoughtful online
  participation}. In \bibinfo{booktitle}{\emph{Proceedings of the SIGCHI
  conference on human factors in computing systems}}.
  \bibinfo{pages}{3401--3410}.
\newblock


\bibitem[\protect\citeauthoryear{Tajfel, Turner, Austin, and Worchel}{Tajfel
  et~al\mbox{.}}{1979}]%
        {tajfel1979integrative}
\bibfield{author}{\bibinfo{person}{Henri Tajfel}, \bibinfo{person}{John~C
  Turner}, \bibinfo{person}{William~G Austin}, {and} \bibinfo{person}{Stephen
  Worchel}.} \bibinfo{year}{1979}\natexlab{}.
\newblock \showarticletitle{An integrative theory of intergroup conflict}.
\newblock \bibinfo{journal}{\emph{Organizational identity: A reader}}
  \bibinfo{volume}{56}, \bibinfo{number}{65} (\bibinfo{year}{1979}),
  \bibinfo{pages}{9780203505984--16}.
\newblock


\bibitem[\protect\citeauthoryear{Tanis and Postmes}{Tanis and Postmes}{2005}]%
        {tanis2005social}
\bibfield{author}{\bibinfo{person}{Martin Tanis} {and} \bibinfo{person}{Tom
  Postmes}.} \bibinfo{year}{2005}\natexlab{}.
\newblock \showarticletitle{A social identity approach to trust: Interpersonal
  perception, group membership and trusting behaviour}.
\newblock \bibinfo{journal}{\emph{European Journal of Social Psychology}}
  \bibinfo{volume}{35}, \bibinfo{number}{3} (\bibinfo{year}{2005}),
  \bibinfo{pages}{413--424}.
\newblock


\bibitem[\protect\citeauthoryear{Towne and Herbsleb}{Towne and
  Herbsleb}{2012}]%
        {towne2012design}
\bibfield{author}{\bibinfo{person}{W~Ben Towne} {and} \bibinfo{person}{James~D
  Herbsleb}.} \bibinfo{year}{2012}\natexlab{}.
\newblock \showarticletitle{Design considerations for online deliberation
  systems}.
\newblock \bibinfo{journal}{\emph{Journal of Information Technology \&
  Politics}} \bibinfo{volume}{9}, \bibinfo{number}{1} (\bibinfo{year}{2012}),
  \bibinfo{pages}{97--115}.
\newblock


\bibitem[\protect\citeauthoryear{Turner, Hogg, Oakes, Reicher, and
  Wetherell}{Turner et~al\mbox{.}}{1987}]%
        {turner1987rediscovering}
\bibfield{author}{\bibinfo{person}{John~C Turner}, \bibinfo{person}{Michael~A
  Hogg}, \bibinfo{person}{Penelope~J Oakes}, \bibinfo{person}{Stephen~D
  Reicher}, {and} \bibinfo{person}{Margaret~S Wetherell}.}
  \bibinfo{year}{1987}\natexlab{}.
\newblock \bibinfo{booktitle}{\emph{Rediscovering the social group: A
  self-categorization theory.}}
\newblock \bibinfo{publisher}{Basil Blackwell}.
\newblock


\bibitem[\protect\citeauthoryear{Valentino, Brader, Groenendyk, Gregorowicz,
  and Hutchings}{Valentino et~al\mbox{.}}{2011}]%
        {valentino2011election}
\bibfield{author}{\bibinfo{person}{Nicholas~A Valentino}, \bibinfo{person}{Ted
  Brader}, \bibinfo{person}{Eric~W Groenendyk}, \bibinfo{person}{Krysha
  Gregorowicz}, {and} \bibinfo{person}{Vincent~L Hutchings}.}
  \bibinfo{year}{2011}\natexlab{}.
\newblock \showarticletitle{Election night’s alright for fighting: The role
  of emotions in political participation}.
\newblock \bibinfo{journal}{\emph{The Journal of Politics}}
  \bibinfo{volume}{73}, \bibinfo{number}{1} (\bibinfo{year}{2011}),
  \bibinfo{pages}{156--170}.
\newblock


\bibitem[\protect\citeauthoryear{Wang}{Wang}{2020}]%
        {wang2020influence}
\bibfield{author}{\bibinfo{person}{Sai Wang}.} \bibinfo{year}{2020}\natexlab{}.
\newblock \showarticletitle{The Influence of Anonymity and Incivility on
  Perceptions of User Comments on News Websites}.
\newblock \bibinfo{journal}{\emph{Mass Communication and Society}}
  \bibinfo{volume}{23}, \bibinfo{number}{6} (\bibinfo{year}{2020}),
  \bibinfo{pages}{912--936}.
\newblock


\bibitem[\protect\citeauthoryear{Wojcieszak and Warner}{Wojcieszak and
  Warner}{2020}]%
        {wojcieszak2020can}
\bibfield{author}{\bibinfo{person}{Magdalena Wojcieszak} {and}
  \bibinfo{person}{Benjamin~R Warner}.} \bibinfo{year}{2020}\natexlab{}.
\newblock \showarticletitle{Can interparty contact reduce affective
  polarization? A systematic test of different forms of intergroup contact}.
\newblock \bibinfo{journal}{\emph{Political Communication}}
  (\bibinfo{year}{2020}), \bibinfo{pages}{1--23}.
\newblock


\bibitem[\protect\citeauthoryear{Yeomans, Minson, Collins, Chen, and
  Gino}{Yeomans et~al\mbox{.}}{2020}]%
        {yeomans2020conversational}
\bibfield{author}{\bibinfo{person}{Michael Yeomans}, \bibinfo{person}{Julia
  Minson}, \bibinfo{person}{Hanne Collins}, \bibinfo{person}{Frances Chen},
  {and} \bibinfo{person}{Francesca Gino}.} \bibinfo{year}{2020}\natexlab{}.
\newblock \showarticletitle{Conversational receptiveness: Improving engagement
  with opposing views}.
\newblock \bibinfo{journal}{\emph{Organizational Behavior and Human Decision
  Processes}} (\bibinfo{year}{2020}).
\newblock


\bibitem[\protect\citeauthoryear{Young}{Young}{2002}]%
        {young2002inclusion}
\bibfield{author}{\bibinfo{person}{Iris~Marion Young}.}
  \bibinfo{year}{2002}\natexlab{}.
\newblock \bibinfo{booktitle}{\emph{Inclusion and democracy}}.
\newblock \bibinfo{publisher}{Oxford University press on demand}.
\newblock


\bibitem[\protect\citeauthoryear{Zheleva and Getoor}{Zheleva and
  Getoor}{2009}]%
        {zheleva2009join}
\bibfield{author}{\bibinfo{person}{Elena Zheleva} {and} \bibinfo{person}{Lise
  Getoor}.} \bibinfo{year}{2009}\natexlab{}.
\newblock \showarticletitle{To join or not to join: the illusion of privacy in
  social networks with mixed public and private user profiles}. In
  \bibinfo{booktitle}{\emph{Proceedings of the 18th international conference on
  World wide web}}. \bibinfo{pages}{531--540}.
\newblock


\end{thebibliography}

\end{document}